\documentclass[english]{article}
\usepackage[T1]{fontenc}
\usepackage[latin9]{inputenc}
\usepackage[letterpaper]{geometry}
\usepackage{float}
\usepackage{amsmath}
\usepackage{graphicx}
\usepackage{color}
\usepackage{epstopdf}

\usepackage[colorlinks, hypertexnames=false]{hyperref}
\hypersetup{linkcolor=blue, citecolor=red} 

\usepackage{adjustbox}
\usepackage{bm}
\usepackage{multirow}
\usepackage{array}
\usepackage{mathrsfs}
\usepackage{amsmath, amssymb, amsthm}
\usepackage{subfigure}
\usepackage{tikz}

\usepackage{fullpage,comment}
\def\d{{\, \rm d}}\newtheorem{prop}{Proposition}

\graphicspath{{./Figures/}}

\usepackage{amsfonts} 
\usepackage{color}
\usepackage{adjustbox}
\definecolor{black}{rgb}{0,0,0}

\definecolor{red}{rgb}{1,0,0}

\definecolor{blue}{rgb}{0,0,1}

\usepackage{multirow}

\makeatother

\usepackage{babel}
\usepackage{authblk}
\title{}

\title{\textbf{}}

\title{An Efficient and Statistically Accurate Lagrangian Data Assimilation Algorithm with Applications to Discrete Element Sea Ice Models}
\author{
	Nan Chen \thanks{Department of Mathematics, University of Wisconsin-Madison, WI, USA.}, \quad
	Shubin Fu \thanks{Department of Mathematics, University of Wisconsin-Madison, WI, USA. Corresponding author: shubinfu89@gmail.com},
	\quad  Georgy E Manucharyan\thanks{School of Oceanography, University of Washington, Seattle, WA 98115, USA. }\;
}
\date{\today}
\begin{document}
	
	\maketitle
	
	\abstract{Lagrangian data assimilation of complex nonlinear turbulent flows is an important but computationally challenging topic. In this article, an efficient data-driven statistically accurate reduced-order modeling algorithm is developed that significantly accelerates the computational efficiency of Lagrangian data assimilation. The  algorithm starts with a Fourier transform of the high-dimensional  flow field, which is followed by an effective model reduction that retains only a small subset of the Fourier coefficients corresponding to the energetic modes. Then a linear stochastic model is developed to approximate the nonlinear dynamics of each Fourier coefficient. Effective additive and multiplicative noise processes are incorporated to characterize the modes that exhibit Gaussian and non-Gaussian statistics, respectively. All the parameters in the reduced order system, including the multiplicative noise coefficients, are determined systematically via closed analytic formulae. These linear stochastic models succeed in forecasting the uncertainty and facilitate an extremely rapid data assimilation scheme. The new Lagrangian data assimilation is then applied to  observations of sea ice floe trajectories that are driven by atmospheric winds and turbulent ocean currents. It is shown that observing only about $30$ non-interacting floes in a $200$km$\times200$km domain is sufficient to recover the key multi-scale features of the ocean currents. The additional observations of the floe angular displacements are found to be suitable supplements to the center-of-mass positions for improving the data assimilation skill. In addition, the observed large and small floes are more useful in recovering the large- and small-scale features of the ocean, respectively. The Fourier domain data assimilation also succeeds in recovering the ocean features in the areas where cloud cover obscures the observations. Finally, the multiplicative noise is shown to be crucial in recovering extreme events.}
	
	~

	\section{Introduction}
	Lagrangian data assimilation is a special but important type of data assimilation problem \cite{apte2013impact, apte2008bayesian, ide2002lagrangian} with wide applications in geophysics, climate science and hydrology \cite{blunden2019look, honnorat2009lagrangian, salman2008using, castellari2001prediction}. Different from Eulerian observations that are at fixed locations, Lagrangian data assimilation exploits the trajectories of moving tracers (e.g., drifters or floaters) as observations to recover the underlying flow field that is often hard to be observed directly. These Lagrangian tracers have particular significance for autonomous data collection in the ocean \cite{griffa2007lagrangian, gould2004argo}.
	
	However, Lagrangian data assimilation faces several computational challenges. First, the underlying flow field that drives the Lagrangian tracers is often high-dimensional with multiscale features, which is due to the strong turbulent nature of the flow field in many geophysical systems \cite{vallis2017atmospheric, pedlosky1987geophysical}. To this end, a high resolution numerical solver is required to not only simulate the key features of the underlying dynamics across different spatiotemporal scales but also guarantee the numerical stability. However, the demand of such a refined numerical scheme brings about a large computational cost when running the model forward at the forecast stage of data assimilation. This is particularly a troublesome issue when the widely-used ensemble data assimilation methods are carried out that require to run the forecast model multiple times in each assimilation cycle.
	Second, despite the Lagrangian observations, the underlying flow field is typically modeled under the Eulerian coordinates. Therefore, it is essential to develop an effective coordinate transformation algorithm beyond the crude interpolations to facilitate the Lagrangian data assimilation.
	Third, the Lagrangian data assimilation is often highly nonlinear \cite{apte2013impact, apte2008bayesian, chen2014information}, which together with the high dimensionality makes it impossible to adopt the exact Bayesian formula to estimate the state variables (unless in very special situations). The intrinsic nonlinearity also requires a careful design of suitable approximate numerical schemes for data assimilation to prevent filter divergence. Similarly, the non-Gaussian feature resulting from strong intermittency of nature is another major challenge that needs to be taken into account. The Lagrangian data assimilation algorithms should thus be able to accurately estimate the states of extreme events and intermittencies associated with the non-Gaussian characteristics.
	During the past two decades, several approximate data assimilation algorithms have been developed \cite{molcard2003assimilation, barreiro2009data, kuznetsov2003method, salman2006method, salman2008using, apte2008bayesian, chen2016model}, which lead to reasonably satisfactory numerical results in certain applications. In particular, the issue of the nonlinear observations can be overcome by augmenting the state variables which also includes the trajectories of Lagrangian tracers \cite{ide2002lagrangian, sun2019lagrangian}.

	The focus of this article is to build an efficient Lagrangian data assimilation algorithm with a systematic reduced order modeling procedure to cope with the high-dimensional complex nonlinear dynamical systems with  multiscale features and non-Gaussian phenomena.  A data-driven Fourier domain data assimilation strategy is developed that aims at significantly reducing the computational cost compared with running the original expensive forecast models at each assimilation cycle. In this new strategy, a Fourier transform is applied to the spatiotemporal patterns associated with the original complex turbulent system, which results in a set of time series of the Fourier coefficients. This is followed by a systematic model reduction in the Fourier domain, maintaining only a small set of the Fourier coefficients corresponding to the energetic modes. Then the complicated dynamics of each Fourier coefficient is effectively approximated by a simple linear stochastic model to advance the computational efficiency. If the long-term statistics of the time series is nearly Gaussian, then a linear model with additive noise (i.e., an Ornstein-Uhlenbeck (OU) process) is adopted as an approximation \cite{gardiner2009stochastic}. Otherwise, a linear stochastic model with multiplicative noise is utilized to characterize the non-Gaussian features.
	One of the advantages of these reduced order linear models is that all the parameters, including the multiplicative noise coefficients, can be systematically determined via closed analytic formulae. Another advantage of the strategy is that the forecast uncertainty due to the nonlinearity between different Fourier modes in the original system is compensated by the stochastic noise in these linear models, which allows each Fourier mode to evolve independently in the forecast stage that significantly reduces the computational cost. These stochastic models can nevertheless provide similar forecast statistics as the original nonlinear model, including the crucial non-Gaussian distributions. Such a statistically accurate forecast is essential to guarantee an accurate data assimilation result. Another key feature of the new strategy is that recovering the Fourier coefficients facilitates the reconstruction of the variables in physical space under the Lagrangian coordinates, which automatically provides an effective coordinate transformation between the Lagrangian floe model and the Eulerian ocean models.
	
\begin{figure}
		\hspace*{-0.0cm}\includegraphics[width=15cm]{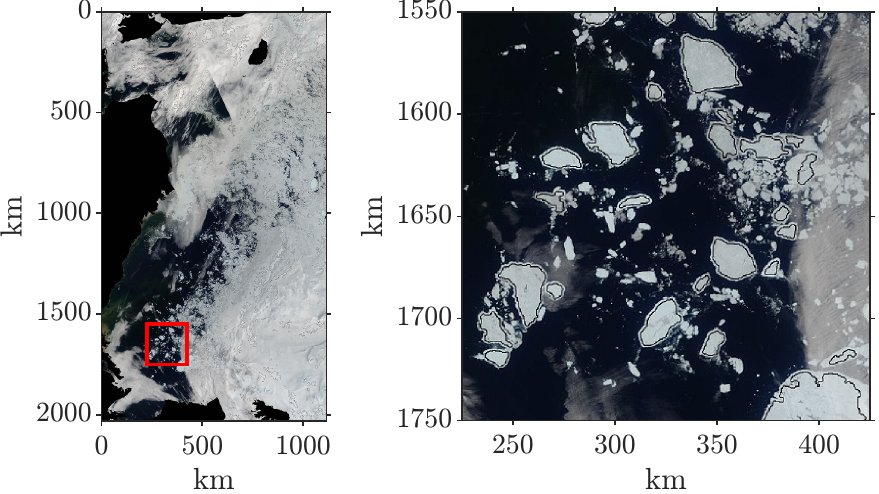}
		\caption{A satellite image of the sea ice floes in the marginal ice zone of the Beaufort sea (north of Alaska) on June 24, 2008. The right panel is an enlarged illustration of the red box area in the left panel. It shows a $200$km$\times200$km domain that contains several sea ice floes. Note that there is a piece of thin cloud cover around the east boundary in the right panel, which blurs the observations of the sea ice floes. The black curves around the floes are added manually from a postprocessing of the figure in order to identify the floes.}\label{Satellite_Image}
	\end{figure}

	The new efficient data assimilation algorithm is then applied to a discrete element sea ice model forced by the  atmosphere and ocean. Sea ice motion is particularly challenging to forecast in marginal ice zones \cite{thomas2017sea,weeks1986growth, zhang2015sea}, where it is not only necessary to consider atmospheric winds but also the sea ice interactions with eddying ocean currents \cite{horvat2016interaction, manucharyan2017submesoscale}. Although at sufficiently large scales the sea ice is widely modeled as a continuum with a given  rheology \cite{hibler1979dynamic, hunke1997elastic, tremblay1997modeling,bouillon2015presentation,rampal2016nextsim}, at scales of the order of 10 km and smaller the sea ice exhibits brittle behavior with individual fragments clearly visible from satellite observations (Figure \ref{Satellite_Image}). For this reason, the discrete element method (DEM) \cite{cundall1979discrete, cundall1988formulation, hart1988formulation} has recently been applied to describing the sea ice dynamics \cite{lindsay2004new, damsgaard2018application, tuhkuri2018review}. The DEM models characterize the motion of each individual sea ice floe under the Lagrangian coordinates, which facilitates the computations compared with the traditional continuum models by avoiding the advective transport scheme and allowing to adaptively change the spatial resolution. The observed sea ice floe trajectories are natural Lagrangian observations that can be used to recover the ocean flow field underneath the sea ice floes, the direct observational data of which is typically hard to obtain.

	In the following, a recently developed DEM sea ice model (SubZero) is utilized to characterize the motion of floes subject to atmospheric winds and eddying ocean currents. The model represents realistic geometric  properties of sea ice floes using non-convex polygons, simulating their physically-consistent nonlinear interactions with the ocean and atmosphere. The ocean model is a two-layer quasi-geostrophic (QG) equation \cite{arbic2004baroclinically} that induces eddies and vortices across different spatial scales \cite{mana2014toward, vallis2017atmospheric}. The parameters of the QG model were calibrated to reproduce the characteristic eddy scales (10 to 50 km) and velocities (5-30 cm/s) in the south-western quarter of the Beaufort Gyre where a large number of ice floe have been detected in its marginal ice zone \cite{lopez2021library}.
	Reanalysis data is adopted for the large-scale atmospheric wind \cite{olauson2018era5}, which is on average one or two orders of magnitude larger than the velocity of the ocean current. These features lead to a high-dimensional multiscale complex turbulent system.
	Only the non-interacting floes are utilized in this study, which avoids dealing with the complicated collision process in data assimilation. It has been shown that these non-interacting floes can be distinguished from the interacting ones in postprocessing the satellite images \cite{lopez2019ice, lopez2021library}.
	One of the primary goals of our Lagrangian data assimilation is to recover the large- and meso-scale features in the ocean, which are important characteristics of ocean turbulence but are not observed directly from the satellites. Another goal is to explore how the number and the size of  sea ice floes in affecting the data assimilation skill, especially for recovering the turbulent ocean eddy field and the extreme events in the atmospheric wind field. 
	Finally, the sea ice floes that are detectable from satellites are of relatively large sizes (about 5-80 km) \cite{lopez2021library} and in addition their observations are often missing due to the presence of clouds \cite{zwally1983antarctic, lopez2019ice}, which are the main differences from traditional Lagrangian observations via drifters or floats. 
	Exploring the Lagrangian data assimilation skill and the associated uncertainties in the presence of the intermittent missing observations is a practically important topic.

	The rest of the article is organized as follows. Section \ref{Sec:Models} describes the  atmosphere-ocean-sea ice system. The new general Lagrangian data assimilation algorithm with the efficient and statistically accurate reduced order forecast models is developed in Section \ref{Sec:DA}. The setup of applying the new Lagrangian data assimilation algorithm to the  atmosphere-ocean-sea ice system is shown in Section \ref{Sec:DA_Setup}. The numerical results are presented in Section \ref{Sec:Tests}. Section \ref{Sec:Discussions} contains discussions for model error, the significance of multiplicative noise and the data assimilation skill with different setups of the model. The article is concluded in Section \ref{Sec:Conclusion}.

	\section{The Coupled Atmosphere-Ocean-Sea Ice Model}\label{Sec:Models}
	We start with introducing the coupled atmosphere-ocean-sea ice model, which  will be used to generate the true signals. The observations of the sea ice floe trajectories will be given by adding certain observational noise to the associated true signals. These true signals will also be used to assess the data assimilation skill.
	
	We postpone the development of the general efficient Lagrangian data assimilation to Section \ref{Sec:DA} since the explanation of the algorithm will be greatly facilitated with this coupled model being served as a concrete example.

	\subsection{The DEM model for the motions of the sea ice floes}
	
	The DEM approach is utilized to describe the motion of the sea ice floes. The floes are modeled by polygons and are treated as rigid bodies. The motion of each floe is characterized by its position (i.e., linear displacement) ${\bf x}=(x,y)$ and its angular displacement $\Omega$ utilizing the following governing equations:
	\begin{equation}\label{sea_ice_basic}
	m\frac{\d^2{\bf x}}{\d t^2}= \iint_{A}{\bf F} \d A + {\bf F}_\text{contact},\qquad\mbox{and}\qquad I\frac{\d^2\Omega}{\d t^2}=\iint_{A}\tau \d A + {\tau}_\text{contact},
	\end{equation}
	where the position of each floe is represented by its center of mass. The second-order time derivative of the linear displacement ${\bf x}$ stands for the acceleration, which is driven by  the contact force with other floes ${\bf F}_\text{contact}$ and the remaining total force $\bf F$ (details will be shown below) integrated over the area of the sea ice floe. Similarly, the acceleration of the angular displacement $\Omega$ is a response of the torque ${\tau}_\text{contact}$ due to the contacting with other floes and the other torque forces $\tau$. Here, $t$ is the time, $m$ is the mass of the floe, $I$ is the moment of inertia and $A$ is the area of the floe.
	By introducing the linear velocity of the floe at its mass center ${\bf v}_{\text{cen}}= ({u}_{\text{cen}}, {v}_{\text{cen}})$ and the angular velocity $\omega$, the model \eqref{sea_ice_basic} can be rewritten as a set of the first-order differential equations,
	\begin{equation}\label{eqfirst}
	\begin{aligned}
	\frac{\d\bf x}{\d t}=&{\bf v}_{\text{cen}},\qquad& m\frac{\d\bf{v}_{\text{cen}}}{\d t}=& \iint_{A}{\bf F} \d A+ {\bf F}_\text{contact}\\
	\frac{\d\Omega}{\d t}=&\omega,\qquad&
	I\frac{\d\omega}{\d t}=&\iint_{A}\tau \d A + {\tau}_\text{contact}.
	\end{aligned}
	\end{equation}
	The total force {\bf F} has four components:
	\begin{equation}\label{floe_force}
	{\bf F}={\bf F}_{\text{ocn}}  +{\bf F}_{\text{atm}}+{\bf F}_{\text{pres}}+{\bf F}_{\text{cor}}
	\end{equation}
	where
	${\bf F}_{\text{ocn}} $ is the drag force induced by the ocean current,
	${\bf F}_{\text{atm}}$ is the drag force induced by the atmospheric wind,
	${\bf F}_{\text{pres}}$ is the forced induced by the tilt of the sea surface height, and
	${\bf F}_{\text{cor}}$ is the Coriolis force.
	Let
	\begin{equation}
	{\bf F}_\text{io}={\bf F}_{\text{ocn}}  +{\bf F}_{\text{atm}}+{\bf F}_{\text{pres}}
	\end{equation}
	be the total force excluding the Coriolis one. Then, the torque $\tau$ is defined as
	\begin{equation}\label{eqt}
	\tau=({\bf r-x})\times  {\bf F}_\text{io}
	\end{equation}
	where $\bf r$ denotes the position vector at a specific point on the floe and `$\times$' is the cross product between two vectors. It is important to note that the ocean forcing of the ice floes nonlinear. The nonlinearity comes from the quadratic drag force ${\bf F}_{\text{ocn}}$, which is given by
	\begin{equation}\label{eqfocn1}
	{\bf F}_{\text{ocn}}=\rho_{\text{ocn}}c_{\text{ocn}}|{\bf v}_{\text{ocn}}-{\bf v}_{\text{ice}}| \bf{R}(\theta)({\bf v}_{\text{ocn}}-{\bf v}_{\text{ice}})
	\end{equation}
	where $\rho_{\text{ocn}}$ is the density of the ocean water, $c_{\text{ocn}}$ is the ocean drag coefficient, $\bf{R}(\theta)$ is the rotation matrix changing the direction of the stress with respect to the velocity difference by a turning angle $\theta$, and ${\bf v}_{\text{ice}}$ is the point-wise ice velocity composed of the translational and rotational velocities.
	Similarly, the drag force ${\bf F}_{\text{atm}}$ induced by the wind is given by
	\begin{equation}\label{eqfatm1}
	{\bf F}_{\text{atm}}=\rho_{\text{atm}}c_{\text{atm}}|{\bf v}_{\text{atm}}|{\bf v}_{\text{atm}}
	\end{equation}
	where $\rho_{\text{atm}}$ is the density of the air, $c_{\text{atm}}$ is ice-wind drag coefficient and ${\bf v}_{\text{atm}}$ is the velocity of the wind. Note that since the ocean current is often of the same order as the sea ice floe velocity, the difference between them, i.e., ${\bf v}_{\text{ocn}}-{\bf v}_{\text{ice}}$, is utilized in \eqref{eqfocn1} to compute the quadratic drag force from the ocean. On the other hand, the atmospheric wind speed is much faster than the sea ice floe motion and therefore the floe velocity is often ignored in \eqref{eqfatm1} \cite{damsgaard2018application, lindsay2004new, rampal2016nextsim}.
	The details of the contact forces are not important here since for data assimilation only observations of non-interacting floes will be utilized. Other details of the floe model are included in the Appendix.

	\subsection{The ocean model}
	The ocean is driven by a two-layer QG equation \cite{arbic2004baroclinically, vallis2017atmospheric}, which is written for potential vorticity anomalies $(q_1, q_2)$ from a pre-defined background state with a mean vertically-sheared flow. The model uses periodic boundary conditions in both the $x$ and $y$ directions. The QG model utilized here is as follows,
	\begin{subequations}\label{eq:qg}
		\begin{align}
		\frac{\partial q_1}{\partial t}+\overline{u_1}\frac{\partial q_1}{\partial x}+\frac{\partial \overline{q_1}}{\partial y}\frac{\partial \psi_1}{\partial x}+J(\psi_1, q_1)= & R_1 \text{div} \big(	\sqrt{\nabla \psi_1\cdot \nabla \psi_1}\nabla \psi_1\big),\label{eq:qg1}\\
		\frac{\partial q_2}{\partial t}+\overline{u_2}\frac{\partial q_2}{\partial x}+\frac{\partial \overline{q_2}}{\partial x}\frac{\partial \psi_2}{\partial x}+J(\psi_2, q_2)= & - R_2\nabla^2\psi_2, \label{eq:qg2}
		\end{align}
	\end{subequations}
	where $\psi_{1,2}$ are the streamfunctions in both layers, the Jacobi is defined as $J(A,B)=\partial A/\partial x\partial B/\partial y-\partial A/\partial y\partial B/\partial x$. The overbars in \eqref{eq:qg} denote the imposed long-term average of the quantities. All the other quantities are anomalies. The variables $u$ and $v$ denote the $x$- and $y$-direction velocities, respectively, while the subscripts $1$ and $2$ denote the upper  and lower layers.  The right-hand-side terms represent the influence of dissipation on the potential vorticity evolution, with the bottom layer having a linear (Ekman-type) drag and the top layer a nonlinear quadratic drag due to sea ice.  Since the ocean eddies evolve much slower than the characteristic timescale needed for an individual flow to pass it, the ocean does not respond to individual sea ice floes but instead to the cumulative impact many passing floes that is represented via the quadratic surface drag with an effective drag coefficient $R_1$.

	The relationship between the streamfunction and the velocity fields is
	\begin{equation}\label{uv}
	(u_1, v_1) =\bigg(-\frac{\partial \psi_1}{\partial y},
	\frac{\partial \psi_1}{\partial x}\bigg)\quad \text{and} \quad
	(u_2, v_2) =\bigg(-\frac{\partial \psi_2}{\partial y},
	\frac{\partial \psi_2}{\partial x}\bigg).
	\end{equation}
	The streamfunctions $\psi_1$ and $\psi_2$ satisfy
	\begin{equation}\label{eq:psi}
	q_1 = \nabla ^2 \psi_1 +\frac{(\psi_2-\psi_1)}{(1+\delta)L^2_d} \quad \text{and} \quad
	q_2 = \nabla ^2 \psi_2 +\frac{\delta(\psi_1-\psi_2)}{(1+\delta)L^2_d}
	\end{equation}
	where $\delta$ is the ratio $H_1/H_2$ of upper-layer to lower-layer depths and $L_d$ is the deformation radius. The second deformation mode radius of about 5.5 km was chosen for $L_d$ as it is more appropriate in describing the length scales of the upper-ocean eddies. Note that the potential vorticity is different from the relative vorticity, which describes the local spinning motion (i.e., the rotation) of a continuum near some point and is defined as
	\begin{equation}\label{eq:vort}
	\xi_1 = \nabla ^2 \psi_1 = \frac{\partial v_1}{\partial x}-\frac{\partial u_1}{\partial y} \quad \text{and} \quad  \xi_2 = \nabla ^2 \psi_2 = \frac{\partial v_2}{\partial x}-\frac{\partial u_2}{\partial y}
	\end{equation}
	Finally, the imposed mean potential vorticity gradients are
	\begin{equation}
	\frac{\partial \overline{q_1}}{\partial y}=
	\frac{(\overline{u_1}-\overline{u_2})}{(1+\delta)L^2_d}
	\quad \text{and} \quad
	\frac{\partial \overline{q_2}}{\partial y}=
	\frac{(\overline{u_2}-\overline{u_1})}{(1+\delta)L^2_d}
	\end{equation}
	The bottom boundary layer thickness $d_{\text{Ekman}}$ determines $R_2$ by $R_2=f_0d_{\text{Ekman}}/(2H_2)$, where $f_0$ is the Coriolis parameter (an f-plane approximation without the beta effect). 

	\subsection{The reanalysis data of the atmosphere}
	The fifth generation ECMWF reanalysis data (ERA5) \cite{olauson2018era5} for the global climate and weather is utilized for describing the atmospheric wind. The box area considered here is 70$^o$N-72.8$^o$N and 136$^o$W-142$^o$W, which is a part of the marginal ice zone in the Beaufort sea where the observations of floe trajectories have been reported \cite{lopez2021library}.
	
	\subsection{Model setups}
	A square domain with a size $200$km$\times200$km is utilized for the study, with double periodic boundary conditions for all the model components. The QG equations were solved using a pseudo-spectral method \cite{arbic2004baroclinically} with a $128\times 128$ spectral modes and a second-order time discretization scheme. The domain size is consistent with the selected box area for the atmospheric reanalysis data.
	
	The linear trends in both the east-west and north-south are removed from the atmospheric data to guarantee the periodicity of the wind. Only the large-scale data of the atmospheric wind is utilized here. The main reason is that the atmospheric wind is more homogeneous in space and the large-scale features are the dominant contributions. This is very different from the ocean field that contains many meso-scale eddies (vortices) at the relatively short length scales of 10-50 km. In addition, the actual observations of the atmospheric wind data are quite sparse in space, which means the information provided by these observations are accurate at only the large scales. To this end, only the $5$ leading Fourier modes are retained in the atmospheric wind velocity data in both $x$ and $y$ directions for the study here. These $5$ modes are $(0,0)$, $(\pm1, 0)$ and $(0,\pm1)$.
	
\begin{figure}
		\hspace*{-0.5cm}\includegraphics[width=18.0cm]{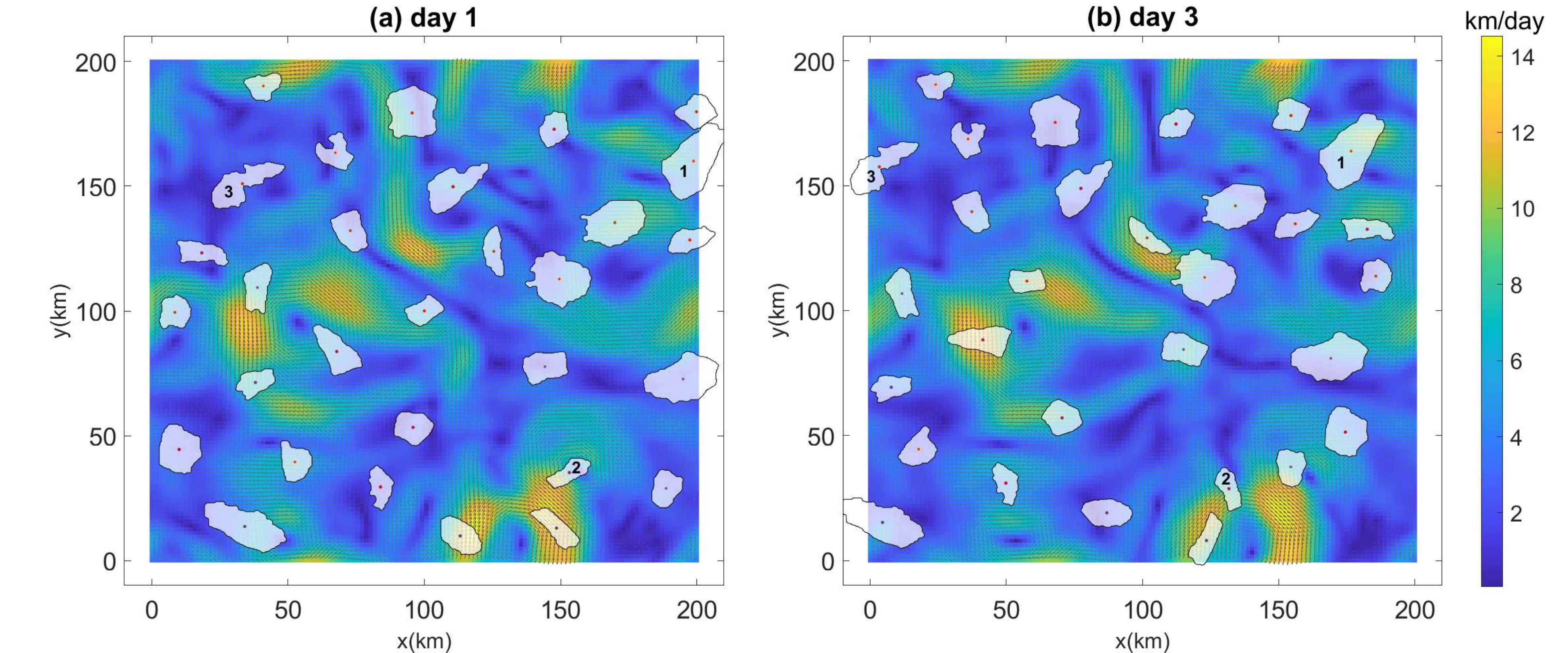}
		\caption{A simulation of the coupled model at day 1 and day 3, where $30$ floes are included. The background contour plot shows the ocean field. Each polygon describes one sea ice floe and the red dot in each floe indicates the center of mass. There are three floes that are marked as Floe \#1, \#2 and \#3.}\label{Flow_Ocean_Illustration}
	\end{figure}

	The sea ice floes were then seeded with random initial locations over the ocean (Fig. \ref{Flow_Ocean_Illustration}), with floe shapes sampled randomly from a library of floe observations over the Beaufort Gyre \cite{lopez2021library}. The floes are forced by the ocean currents and atmospheric winds but they do not affect the oceanic and atmospheric dynamics.
	
	The model parameters are summarized in Table \ref{Table_Parameters}. The coupled atmosphere and ocean system is multiscale in both time and space. In fact, the atmospheric wind velocity is typically $8$m/s to $10$m/s (corresponding to $800$km/day), which is much faster than the ocean current speed that is roughly $0.1$m/s (corresponding to $10$km/day). On the other hand, the atmospheric wind changes rapidly in time while the temporal variation of the ocean current is much more slowly.
	
	\begin{table}[t]
		\caption{The model parameters setup.}\label{Table_Parameters}
		\begin{center}
			\begin{tabular}{lc}
				\hline\hline
				Physical values \\
				\hline
				Domain size & $ 200$km$\times200$km  \\
				Mesh size&$1.5625$km$\times$ $1.5625$km\\
				Time step &$1/384$days\\
				Ocean density & $\rho_{\text{ocn}} =1027$kg/m$^3$  \\
				Ice density &$\rho_{\text{ice}}=920$kg/m$^3$ \\
				Air density & $\rho_{\text{atm}}=1.2$kg/m$^3$  \\
				Ocean drag coefficient & $c_{\text{ocn}}=5.5\times 10^{-3}$\\
				Atmosphere drag coefficient & $c_{\text{atm}}=1.6\times 10^{-3}$\\
				Coriolis coefficient & $f_c=1.4\times 10^{-4}$\\
				Top layer mean ocean velocity & $\overline{u_1}=2.58$km/day\\
				Bottom layer mean ocean velocity & $\overline{u_2}=1.032$km/day\\
				Top layer mean potential vorticity&$\frac{\partial \overline{q_1}}{\partial y}=0.0265$km$^{-1}$day$^{-1}$\\
				Bottom layer mean potential vorticity&$\frac{\partial \overline{q_2}}{\partial x}=-0.0212$km$^{-1}$day$^{-1}$\\
				Coriolis parameter&$f_c=12$day$^{-1}$\\
				Coupling parameter &$R_1=6.9\times10^{-5}$km$^{-1}$\\
				Decay rate of the barotropic mode&$R_2=1$day$^{-1}$\\
				Deformation radius	&$L_d=5.7$km\\
				Ratio of upper-to lower-layer depth&$\delta=0.8$\\
				Turning angle of the ocean & $\theta=\pi/9$\\
				Floe thickness &$h=1$m\\
				\hline
			\end{tabular}
		\end{center}
	\end{table}
	
\begin{figure}
		\hspace*{-0.5cm}\includegraphics[width=18.0cm]{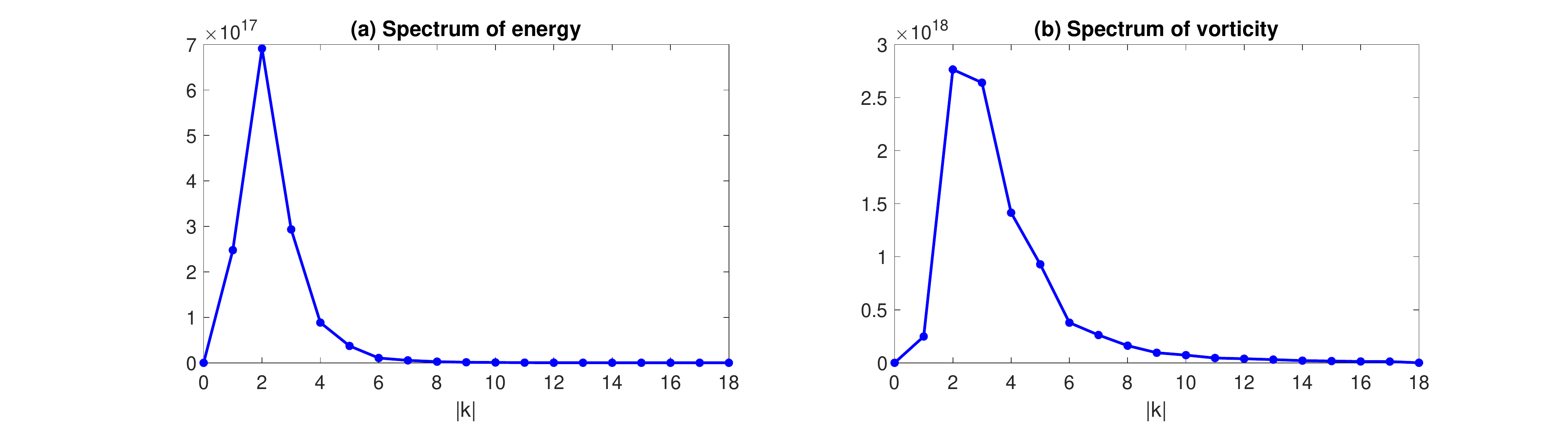}
		\caption{Spectrum of the energy and the spectrum of the vorticity associated with the ocean QG model in the upper layer. Note that the spectrum value corresponding to $|\mathbf{k}|$ includes all those modes that lie inside the interval $[|\mathbf{k}|,|\mathbf{k}|+1)$. The energy in the zeroth mode is zero.}\label{Spectrums}
	\end{figure}

	Figure \ref{Spectrums} shows the energy and the vorticity spectra corresponding to the upper (surface) layer of the ocean QG model. Here, the spectrum is shown as a function of the absolute value of the wavenumber $|\mathbf{k}| = \sqrt{k_1^2+k_2^2}$. The spectrum of $\mathbf{k}$ stands for the total energy or vorticity summing over the modes inside the interval $[|\mathbf{k}|,|\mathbf{k}|+1)$. The energy spectrum peaks at $|\mathbf{k}|=2$ and the ocean is energetic up to at least $|\mathbf{k}|=6$. The vorticity has a wider spectrum with a non-negligible contribution up to at least  $|\mathbf{k}|=13$. Note that the ocean model here focuses on representing its internal variabilities. In other words, the spectrum does not peak at the largest spatial scale $|\mathbf{k}|=0$, at which the ocean and the sea ice floes are mostly driven by the atmospheric wind resulting in a motion that is spatially homogeneous. At the second largest spatial scale, $|\mathbf{k}|=1$, both the atmospheric wind and the ocean current have contributions to the floe motion but the wind lies in a much faster time scale compared with the ocean current. Therefore, the contributions from the atmosphere and the ocean can be distinguished.

	Figure \ref{Flow_Ocean_Illustration} shows a simulation of the coupled model. The background contour plot indicates the ocean field. Each polygon in white color describes one sea ice floe, where the red dot represents the center of mass. The left panel shows the floes and the ocean field at day 1 while the right panel shows them at day 3. It is clear that all the floes have a tendency to move towards the west. This is mainly due to the atmospheric wind, since for example the ocean current around the floe \#3 is very weak.  On the other hand, the ocean vorticity can cause the rotation of the sea ice floes. For example, a strong ocean vortex is observed underneath floe \#2, which leads to an anti-clockwise rotation of the floe for about 90$^o$.

	\section{An Efficient and Statistically Accurate Data Assimilation Method}\label{Sec:DA}
	\subsection{Motivation}
	Data assimilation algorithms contain two steps in each assimilation cycle: 1) forecast, and 2) analysis. The state variables of the coupled system are:
	\begin{enumerate}
		\item [(a). ]the locations $\bf x$ and the angular displacements $\Omega$ of the floes,
		\item [(b). ]the velocities ${\bf v}_\text{cen}$ and the angular velocities $\omega$ of the floes, and
		\item [(c). ]the ocean and the atmosphere flow fields.
	\end{enumerate}
	The observations are only the locations $\bf x$ and the angular displacements $\Omega$ of the floes.
	
	Denote $L$ the number of the floes, and $D_o$ and $D_a$ the dimension of the ocean and atmosphere models, respectively. The total dimension of the forecast model, assuming to use the perfect system, will be $3L$, $3L$, and $D_o + D_a$ for parts (a), (b) and (c), respectively. The number of the floes in the standard test will be $L=30$ while the dimension of the ocean will be $D_o=128^2\times 2\approx30,000$. These values indicate that the dominant part of the computational cost in the forecast step is to run the ocean model. In fact, a single run of the QG model with a spatial resolution of $D_o$ is already computationally very expensive, let alone running a number of ensembles. In addition, the computational cost here mainly comes from the forecast instead of the analysis step. Thus, developing efficient approximate models for the full ocean system is crucial for advancing the data assimilation. On the other hand, the atmospheric model here is unknown, which implies building simple data-driven models to describe the atmospheric wind is also crucial for  effective data assimilation.
	
	\subsection{Overview}
\begin{figure}
		\hspace*{-0.5cm}\includegraphics[width=18.0cm]{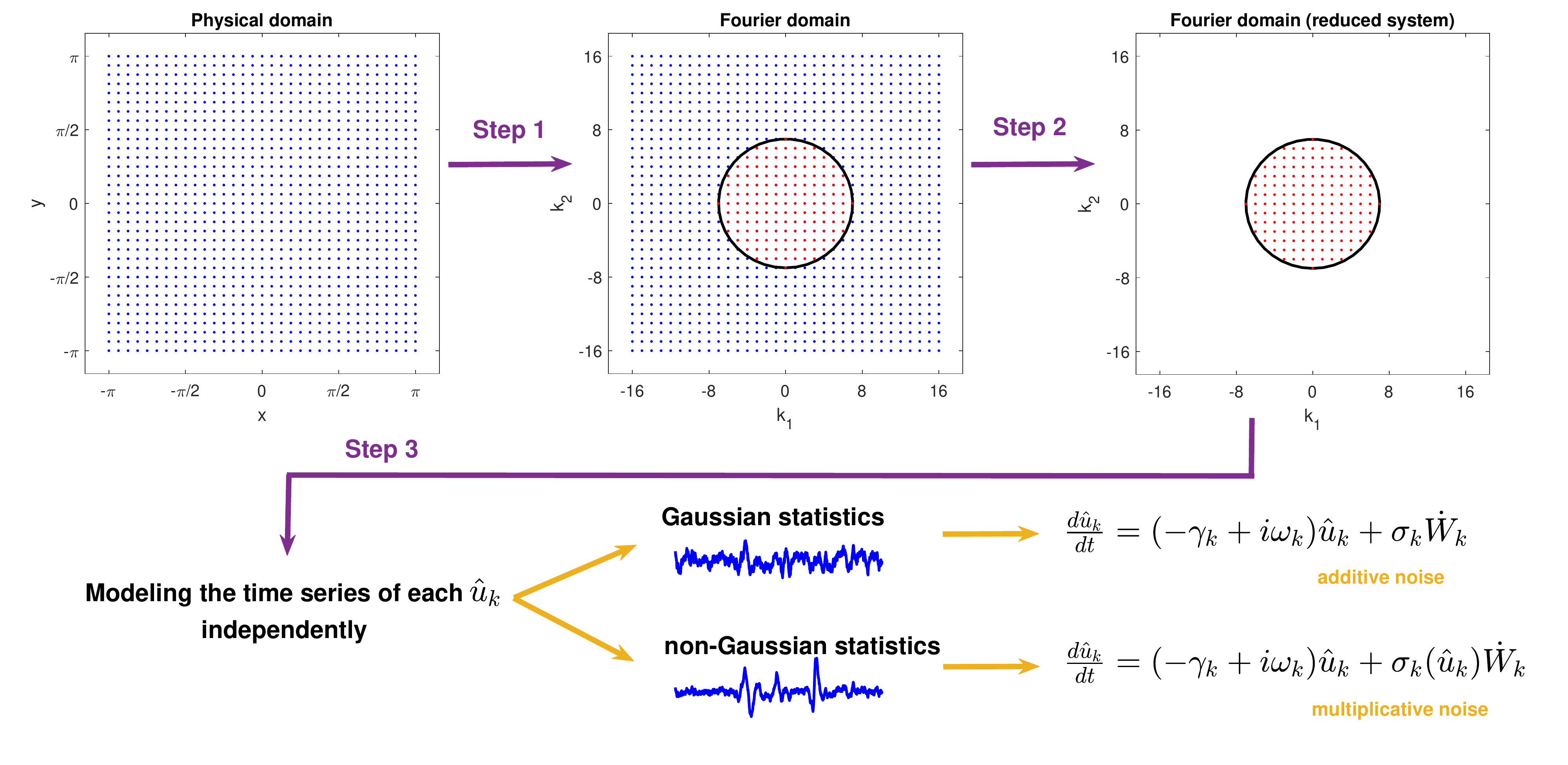}
		\caption{A schematic illustration of the Fourier domain filtering strategy with the efficient and statistically accurate stochastic models for the time evolution of each Fourier coefficient. For the simplicity of illustration, the real-valued mode $\mathbf{k}=(0,0)$ is not included here. The ranges of $k_1$ and $k_2$ adopted here are also simply for the illustration purpose. For the actual ocean model used in this study, the original meshgrid for the QG model is $128\times 128\times 2$ while the reduced order model contains only the modes $|\mathbf{k}|<17$ in the upper layer ocean, which is only about $1/150$ of the total degrees of freedom in the original models.}\label{Illustration}
	\end{figure}

	The localized ensemble transform Kalman filter (LETKF) \cite{ott2004local, hunt2007efficient} is utilized as the basic data assimilation scheme. The localization mitigates the sampling errors that often induce the erroneous spurious spatial correlations and thus allows to use only a small number of the ensembles in the data assimilation procedure. The following procedures are adopted for effectively approximating the underlying flow field (e.g., ocean and atmosphere in the coupled model here) in the new Lagrangian data assimilation algorithm to reduce the computational cost:
	\begin{enumerate}
		\item Transforming the state variables from the physical space to the Fourier space.
		\item Model reduction by retaining only the time series of the energetic Fourier modes.
		\item Developing efficient and statistically accurate stochastic models for approximating the time evolution of each Fourier coefficient retained in Step 2.
	\end{enumerate}
	A schematic illustration of the data assimilation with the reduced order modeling strategy is shown in Figure \ref{Illustration}, where the details will be discussed in the following subsections.

	\subsection{Fourier domain data assimilation}
	The first step towards the development of an efficient approximate forecast model for the ocean component is to transform the state variables from the physical space to the Fourier space \cite{majda2012filtering}. Since many Lagrangian data assimilation problems focus on the center of the ocean, applying the Fourier transform is natural.
	The motivation of the Fourier domain data assimilation is from the energy and vorticity spectrums in Figure \ref{Spectrums}, which shows that the energetic modes in terms of both the energy and the vorticity lie within a certain spectrum band. If the approximate forecast model contains only these modes, then the dimension of the forecast model can be significantly reduced compared with the full ocean QG model. Note that solving a QG model with a spatial resolution that is much coarser than the original one, say $16\times 16\times 2$, will completely change the dynamical features (and the solution in fact blows up in the setup here). Therefore, new strategies are required for the development of effective reduced order models to approximate the  forecast of these  Fourier coefficients of the ocean model. It is also worthwhile to note that the coordinate transform between the Lagrangian floe model and the Eulerian ocean model becomes much simpler when the ocean is represented under the Fourier bases.
	
	Despite being difficult to develop physically consistent simplified models, it suffices to build an approximate system that characterizes the correct uncertainty propagation of the original model for the purpose of skillful data assimilation. This is because the outcome of the forecast step by running the model forward is a probability density function (PDF). If the approximate model, regardless of its exact physics, can generate a forecast PDF that is close to the truth, then the associated data assimilation results will be similar as well. Different from the state variables in the physical space, where the spatial correlation is strong between nearby grid points, the correlation between different Fourier coefficients is often much weaker. Therefore, the Fourier domain data assimilation allows to the development of a forecast system that involves modeling and running forward different Fourier coefficients independently, which facilitates the computational efficiency. In addition, only the energetic ones will be included in the approximate model, which will further reduce the computational cost by a significant amount.
	
	Next, simple linear stochastic models are developed to approximate the time evolution of these Fourier coefficients (see Section \ref{Sec:linearStochasticModels} for details). Depending on the Gaussian or non-Gaussian statistics of each Fourier mode, either a very simple additive noise or a systematic multiplicative noise determined by closed analytic formulae will be adopted in the associated linear stochastic model. One of the key advantages of such a linear stochastic system is that the forecast uncertainty due to the complicated nonlinear interactions between different Fourier modes is effectively characterized by the stochastic noise. Despite having a completely different physics from the truth, these linear stochastic models succeed in capturing the uncertainty in the forecast step, at least for the marginal distributions of each Fourier coefficient. The calibration of the stochastic models here will be based on the true signal of the ocean generated from the QG model, which is the perfect model in this study. In practice, the perfect model is unknown. Nevertheless, any sophisticated nonlinear physical model in hand can be used to calibrate the reduced order stochastic system.
	
	According to Figure \ref{Spectrums}, the modes to be included in the reduced order system are those whose Fourier wavenumber $|\textbf{k}|< 17$, which implies that there are in total only about $200$ modes in the reduced order model. This is much fewer than the degree of freedom of the original QG model, which is around $30,000$. Note that, in principle, aliasing should be taken into account if the data assimilation contains only a subset of the state variables. Nevertheless, since the spectrums of both the energy and the vorticity are nearly zero for $|\textbf{k}| \geq 17$, which means the associated aliasing error is tiny, those modes are simply ignored in data assimilation. Finally, similar linear stochastic models are applied to model each Fourier coefficient of the atmospheric wind.
	
	Below, the notation $\hat{u}_{\mathbf{k}}$ is utilized to represent each Fourier mode for both the atmosphere and ocean but describes different physical variables. For the ocean part, each $\hat{u}_{\mathbf{k}}$ is one Fourier coefficient of the upper layer streamfunction $\psi_1$, which is the variable that couples with the sea ice. Since ocean is incompressible, the velocities in the $x$ and $y$ directions of the upper layer ocean are uniquely determined by the streamfunction \eqref{uv}. Note that although $\psi_2$ appears in the original QG model, it is not directly used in the reduced order system. On the other hand, the atmospheric wind is compressible. Therefore,  the wind velocities in either the $x$ or the $y$ directions need to be modeled separately, the equations of which are both denoted by $\hat{u}_{\mathbf{k}}$.
	
	\subsection{Efficient and statistically accurate linear stochastic models for the time evolution of each Fourier coefficient}\label{Sec:linearStochasticModels}
	
	A complex linear stochastic model is utilized to approximate the time evolution of each Fourier coefficient $\hat{u}_{\mathbf{k}}$ associated with the ocean and atmosphere fields for ${\mathbf{k}}\neq\bf 0$,
	\begin{equation}\label{eq:ocnmode}
	\frac{\d\hat{u}_{\mathbf{k}}}{\d t} = (-\gamma_{\mathbf{k}}+i\omega_{\mathbf{k}})\hat{u}_{\mathbf{k}}+\sigma_{\mathbf{k}} \dot{W}_{\mathbf{k}},
	\end{equation}
	where the real-valued parameters $\gamma_{\mathbf{k}}$, $\omega_{\mathbf{k}}$ and $\sigma_{\mathbf{k}}$ are damping, oscillation and noise coefficients, respectively, while $\dot{W}_{\mathbf{k}}$ is a complex white noise source and $i$ is the imaginary unit. Note that since the ocean model is an anomaly model and the mean atmospheric wind is only contained in the zeroth mode, the mean states of $\hat{u}_{\mathbf{k}}$ is zero for $\bf k\neq \bf 0$. Thus, there is no constant forcing term in \eqref{eq:ocnmode} in such a situation. The mode ${\mathbf{k}}=\mathbf{0}$ for the atmospheric wind will be discussed at the end of this subsection. The damping and oscillation parameters $\gamma_{\mathbf{k}}$ and $\omega_{\mathbf{k}}$ in \eqref{eq:ocnmode} are always assumed to be constants. On the other hand, depending on the long-term statistics, i.e., the PDF, of $\hat{u}_{\mathbf{k}}$, constant or state-dependent noise coefficients $\sigma_{\mathbf{k}}$ will be adopted. \medskip
	
	\noindent\textbf{Case I: Gaussian long-term statistics of $\hat{u}_{\mathbf{k}}$.}\\
	Assume for now the long-term statistics of $\hat{u}_\mathbf{k}$ is Gaussian (or nearly Gaussian in practice). Then a constant $\sigma_{\mathbf{k}}$ is utilized as the noise coefficient in \eqref{eq:ocnmode}. The three parameters $\gamma_{\mathbf{k}}$, $\omega_{\mathbf{k}}$ and $\sigma_{\mathbf{k}}$ can be determined by matching the long-term statistics of \eqref{eq:ocnmode} with the actual time series of the associated Fourier coefficient \cite{majda2012filtering}. The following proposition provides the formulae of obtaining these three parameters.
	
	\begin{prop}\label{prop:ou}
		Assume $\sigma_\mathbf{k}$ is a constant. Then the three parameters $\gamma_\mathbf{k}$, $\omega_\mathbf{k}$ and $\sigma_\mathbf{k}$ in \eqref{eq:ocnmode} are determined utilizing the following formulae
		\begin{equation}\label{OU_Parameters}
		\gamma_\mathbf{k}=\frac{T_\mathbf{k}}{T_\mathbf{k}^2+\theta_\mathbf{k}^2},\qquad
		\omega_\mathbf{k}=\frac{\theta_\mathbf{k}}{T_\mathbf{k}^2+\theta_\mathbf{k}^2}\qquad\mbox{and}\qquad
		\sigma_\mathbf{k}=\sqrt{\frac{2E_\mathbf{k}T_\mathbf{k}}{T_\mathbf{k}^2+\theta_\mathbf{k}^2}}.
		\end{equation}
		where $E_\mathbf{k}$ is the equilibrium variance of $\hat{u}_\mathbf{k}$, namely
		\begin{equation}
		E_\mathbf{k}\equiv \text{Var}(\hat{u}_\mathbf{k})=\overline{|\hat{u}_\mathbf{k}(t)-\overline{\hat{u}}_\mathbf{k}|^2}
		\end{equation}
		with $\overline{\hat{u}}_\mathbf{k}$ being the long-term mean of $\hat{u}_\mathbf{k}$. The values $T_\mathbf{k}$ and $\theta_\mathbf{k}$ are associated with integration of the autocorrelation function (ACF), namely the decorrelation time, of $\hat{u}_\mathbf{k}(t)$
		\begin{equation}
		\int_{0}^{\infty}R_{\hat{u}_\mathbf{k}}(\tau)d\tau=T_\mathbf{k}-i\theta_\mathbf{k}
		\end{equation}
		where the  ACF is given by
		\begin{equation}
		R_{\hat{u}_\mathbf{k}}(\tau)= \frac{\overline{(\hat{u}_\mathbf{k}(t)-\overline{\hat{u}}_\mathbf{k})(\hat{u}_\mathbf{k}(t+\tau)-\overline{\hat{u}}_\mathbf{k})^*}}{\text{Var}(\hat{u}_\mathbf{k})}.
		\end{equation}
	\end{prop}
	The proofs of this and the following propositions are included in the Appendix.\medskip
	
	\noindent\textbf{Case II: non-Gaussian long-term statistics of $\hat{u}_{\mathbf{k}}$.}\\
	Next, consider the situation that the long-term statistics of the Fourier coefficient $\hat{u}_\mathbf{k}$ is non-Gaussian. To characterize such a non-Gaussian feature, a linear model with a state-dependent noise coefficient $\sigma_{\mathbf{k}}(\hat{u}_\mathbf{k})$ in \eqref{eq:ocnmode} is utilized as an approximate model \cite{averina1988numerical},
	\begin{equation}\label{eq:ocnmode2}
	\frac{\d\hat{u}_{\mathbf{k}}}{\d t} = (-\gamma_{\mathbf{k}}+i\omega_{\mathbf{k}})\hat{u}_{\mathbf{k}}+\sigma_{\mathbf{k}}(\hat{u}_\mathbf{k}) \dot{W}_{\mathbf{k}},
	\end{equation}
	For the convenience of discussion, the model \eqref{eq:ocnmode} for $\bf k\neq \bf 0$ is rewritten into the following two-dimensional form,
	\begin{equation}\label{2D_model}
	\begin{split}
	\frac{\d \hat{u}_{\mathbf{k}, 1}}{\d t} &= -\gamma_{\mathbf{k}} \hat{u}_{\mathbf{k}, 1} - \omega_{\mathbf{k}} \hat{u}_{\mathbf{k}, 2}  + \sigma_{\mathbf{k},1}(\hat{u}_{\mathbf{k}, 1},\hat{u}_{\mathbf{k}, 2}) \dot{W}_{\mathbf{k}, 1},\\
	\frac{\d \hat{u}_{\mathbf{k}, 2}}{\d t} &=  - \gamma_{\mathbf{k}} \hat{u}_{\mathbf{k}, 2} + \omega_{\mathbf{k}} \hat{u}_{\mathbf{k}, 1} + \sigma_{\mathbf{k},2}(\hat{u}_{\mathbf{k}, 1},\hat{u}_{\mathbf{k}, 2}) \dot{W}_{\mathbf{k}, 2},\\
	\end{split}
	\end{equation}
	where the real-valued variables $\hat{u}_{\mathbf{k}, 1}$ and $\hat{u}_{\mathbf{k}, 2}$ are the real and imaginary parts of the complex variable $\hat{u}_{\mathbf{k}}$.
	The damping and oscillation coefficients $\gamma_{\mathbf{k}}$ and $\omega_{\mathbf{k}}$ are determined in the same way as those in the Gaussian case \eqref{OU_Parameters}.
	Despite the linear dynamics, the state-dependent noise coefficients $\sigma_{\mathbf{k},1}(\hat{u}_{\mathbf{k}, 1},\hat{u}_{\mathbf{k}, 2})$ and $\sigma_{\mathbf{k},2}(\hat{u}_{\mathbf{k}, 1},\hat{u}_{\mathbf{k}, 2})$ are included to capture the non-Gaussian features of the time series.

	Let $p(\hat{u}_{\mathbf{k}, 1},\hat{u}_{\mathbf{k}, 2})$ be the stationary PDF associated with the system \eqref{2D_model}. The following proposition provides one solution of the multiplicative noise coefficients $\sigma_{\mathbf{k},1}(\hat{u}_{\mathbf{k}, 1},\hat{u}_{\mathbf{k}, 2})$ and $\sigma_{\mathbf{k},2}(\hat{u}_{\mathbf{k}, 1},\hat{u}_{\mathbf{k}, 2})$.
	\begin{prop}\label{Prop_nonG1}
		Given a time series  $\hat{u}_{\bf k} := \hat{u}_{\mathbf{k}, 1} + i \hat{u}_{\mathbf{k}, 2}$ of a Fourier coefficient with both $\hat{u}_{\mathbf{k}, 1}$ and $\hat{u}_{\mathbf{k}, 2}$ being real-valued components. Assume the two constants $\gamma_{\mathbf{k}}$ and $\omega_{\mathbf{k}}$ in \eqref{2D_model} have been determined by matching the decorrelation time of the model with the observational time series utilizing \eqref{OU_Parameters}. Then the multiplicative noise coefficients $\sigma_{\mathbf{k},1}(\hat{u}_{\mathbf{k}, 1},\hat{u}_{\mathbf{k}, 2})$ and $\sigma_{\mathbf{k},2}(\hat{u}_{\mathbf{k}, 1},\hat{u}_{\mathbf{k}, 2})$ can be determined via the following formulae,
		\begin{subequations}\label{Multiplicative_noise_2states}
			\begin{align}
			\sigma^2_{\mathbf{k},1}(\hat{u}_{\mathbf{k}, 1},\hat{u}_{\mathbf{k}, 2}) &= \frac{2}{p(\hat{u}_{\mathbf{k}, 1},\hat{u}_{\mathbf{k}, 2})}\int_{-\infty}^{\hat{u}_{\mathbf{k}, 1}} (-\gamma_{\mathbf{k}} s - \omega_{\mathbf{k}} \hat{u}_{\mathbf{k}, 2}) p(s, \hat{u}_{\mathbf{k}, 2}) \d s, \label{Multiplicative_noise_2states_1}\\
			\sigma^2_{\mathbf{k},2}(\hat{u}_{\mathbf{k}, 1},\hat{u}_{\mathbf{k}, 2}) &= \frac{2}{p(\hat{u}_{\mathbf{k}, 1},\hat{u}_{\mathbf{k}, 2})}\int_{-\infty}^{\hat{u}_{\mathbf{k}, 2}} (- \gamma_{\mathbf{k}} s + \omega_{\mathbf{k}} \hat{u}_{\mathbf{k}, 1} ) p(\hat{u}_{\mathbf{k}, 1},s) \d s. \label{Multiplicative_noise_2states_2}
			\end{align}
		\end{subequations}
	\end{prop}
	
	In practice, a simplification can be made by assuming $\sigma_{\mathbf{k},1}:=\sigma_{\mathbf{k},1}(\hat{u}_{\mathbf{k},1})$ is only a function of $\hat{u}_{\mathbf{k},1}$ and $\sigma_{\mathbf{k},2}:=\sigma_{\mathbf{k},2}(\hat{u}_{\mathbf{k},2})$ is only a function of $\hat{u}_{\mathbf{k},2}$. These assumptions facilitate the calculations of  approximate solutions of the multiplicative noise coefficients $\sigma_{\mathbf{k},1}$ and $\sigma_{\mathbf{k},2}$. Further denote by $p_1(\hat{u}_{\mathbf{k},1})$ and $p_2(\hat{u}_{\mathbf{k},2})$  the marginal distributions of  $p(\hat{u}_{\mathbf{k},1},\hat{u}_{\mathbf{k},2})$.
	\begin{prop}\label{Prop_nonG2}
		Assume  $\sigma_{\mathbf{k},1}:=\sigma_{\mathbf{k},1}(\hat{u}_{\mathbf{k},1})$ is only a function of $\hat{u}_{\mathbf{k},1}$ and $\sigma_{\mathbf{k},2}:=\sigma_{\mathbf{k},2}(\hat{u}_{\mathbf{k},2})$ is only a function of $\hat{u}_{\mathbf{k},2}$, then  approximate solutions to these multiplicative noise coefficients are as follows,
		\begin{equation}\label{Multiplicative_noise_1states}
		\begin{split}
		\sigma_{\mathbf{k},1}^2(\hat{u}_{\mathbf{k}, 1}) &= \frac{-2\gamma_{\mathbf{k}}}{p_1(\hat{u}_{\mathbf{k}, 1})}\int_{-\infty}^{\hat{u}_{\mathbf{k}, 1}} s p_1(s) \d s\\
		\sigma_{\mathbf{k},2}^2(\hat{u}_{\mathbf{k}, 2}) &= \frac{-2\gamma_{\mathbf{k}}}{p_2(\hat{u}_{\mathbf{k}, 2})}\int_{-\infty}^{\hat{u}_{\mathbf{k}, 2}} s p_2(s) \d s.
		\end{split}
		\end{equation}
	\end{prop}
	
	Finally, for mode $\mathbf{k}=(0,0)$ of the atmospheric wind, the following real-valued linear stochastic model is utilized as an approximate model,
	\begin{equation}\label{eq:ocnmode3}
	\frac{\d\hat{u}_{\mathbf{k}}}{\d t} = -\gamma_{\mathbf{k}}\hat{u}_{\mathbf{k}} + f_\mathbf{k}+\sigma_{\mathbf{k}}(\hat{u}_{\mathbf{k}}) \dot{W}_{\mathbf{k}},
	\end{equation}
	where the state variable $\hat{u}_{\mathbf{k}}$, the white noise $\dot{W}_{\mathbf{k}}$ and the parameters are all real-valued for this special mode $\mathbf{k}=(0,0)$. Following the above discussions, the two constant parameters $\gamma_{\mathbf{k}}$ and $f_{\mathbf{k}}$ as well as the multiplicative noise coefficient $\sigma_{\mathbf{k}}(\hat{u}_{\mathbf{k}})$ can be determined as follows,
	\begin{prop}\label{Prop_nonG3}
		The two constant parameters in \eqref{eq:ocnmode3} for mode $\mathbf{k}=(0,0)$ are given by
		\begin{equation}
		\gamma_\mathbf{k}=\frac{1}{T_\mathbf{k}}\qquad\mbox{and}\qquad f_{\mathbf{k}} = \frac{\overline{\hat{u}}_{\mathbf{k}}}{T_\mathbf{k}}.
		\end{equation}
		where $\overline{\hat{u}}_{\mathbf{k}}$ is the long-term mean state of $\hat{u}_{\mathbf{k}}$.
		The multiplicative noise coefficient $\sigma_{\mathbf{k}}(\hat{u}_{\mathbf{k}})$ is solved via
		\begin{equation}
		\sigma^2_{\mathbf{k}}(\hat{u}_{\mathbf{k}}) = \frac{-2\gamma_{\mathbf{k}}}{p(\hat{u}_{\mathbf{k}})}\int_{-\infty}^{\hat{u}_{\mathbf{k}}} \left(s-\frac{f_{\mathbf{k}}}{\gamma_{\mathbf{k}}}\right) p(s) \d s
		\end{equation}
		where $p(\hat{u}_{\mathbf{k}})$ is the equilibrium distribution of $\hat{u}_{\mathbf{k}}$.
	\end{prop}

\begin{figure}
		\hspace*{-0.5cm}\includegraphics[width=18.0cm]{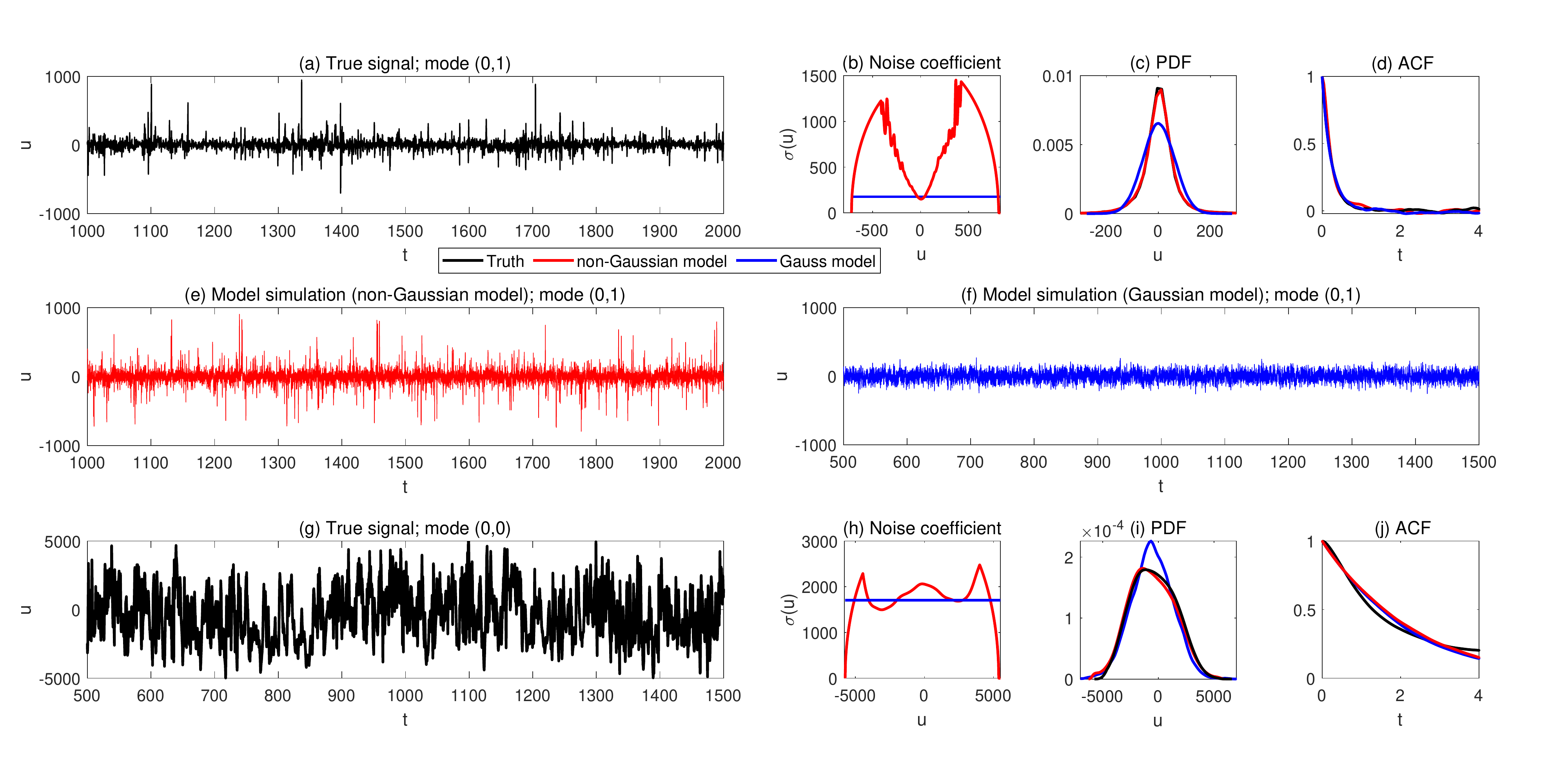}
		\caption{Comparison of the true signal of the atmospheric wind with both the Gaussian and non-Gaussian approximate models. Panel (a): the true signal of the Fourier coefficient $\mathbf{k}=(0,1)$; only the real part is shown. Panel (b): comparing the multiplicative noise coefficient $\sigma({\hat{u}_\mathbf{k}})$ in \eqref{eq:ocnmode2} and the additive noise coefficient $\sigma$ in \eqref{eq:ocnmode} for the two linear models that fit the true signal in Panel (a). Panels (c)--(d): the PDFs and ACFs of the truth, the linear models with multiplicative noise and additive noise. Panels (e): a random realization of the linear model \eqref{eq:ocnmode2} with calibrated multiplicative noise. Panels (f): a random realization of the linear model \eqref{eq:ocnmode} with calibrated additive noise. Panels (g)--(j): similar to Panels (a)--(d) but for the Fourier coefficient $\mathbf{k}=(0,0)$. Note that the time series of the Fourier coefficient $\mathbf{k}=(0,1)$ is strongly non-Gaussian with fat tails while that of the Fourier coefficient $\mathbf{k}=(0,0)$ is strongly sub-Gaussian. }\label{Time_Series_Fits}
	\end{figure}

	Figure \ref{Time_Series_Fits} shows the results of approximating two of the atmospheric modes using the linear stochastic models. Panel (a) shows the observed signal of mode $\mathbf{k}=(0,1)$, which is highly intermittent with a fat-tailed PDF (Panel (c)). Some  non-Gaussian statistics have been pointed out in previous work \cite{qi2016low}. If a linear model with additive noise \eqref{eq:ocnmode} is adopted for approximation, then even with the optimal parameters \eqref{OU_Parameters} the model fails to generate the observed extreme events and the fat-tailed PDF. See the blue curves in Panels (c) and (f). In contrast, the linear model with multiplicative noise is able to capture both the dynamical and statistical features of nature. Specifically, the observed fat-tailed PDF, the ACF and the intermittent trajectories are all recovered by the linear model with multiplicative noise. See the red curves in Panels (c)--(e). Note that the multiplicative noise coefficients are determined by only the approximation formulae in  \eqref{Multiplicative_noise_1states}. Panel (b) illustrates the multiplicative noise (red curve), which is very different from a constant, indicating the necessity of using a state-dependent description of the noise coefficient. Panel (d) indicates that the linear model also reproduces the ACF of the truth, which is as a result of the designing of the model calibration. Recovering the ACF is important for the approximate model to capture the time evolution of the uncertainty in the perfect system.
	Similarly, Panels (g)--(j) show the true trajectory and the statistics associated with  truth and the linear model with multiplicative noise for mode $\mathbf{k}=(0,0)$. Different from the highly intermittent mode $\mathbf{k}=(0,1)$, the statistics of $\mathbf{k}=(0,0)$ is sub-Gaussian with a kurtosis Kurt$=2.28$ that is much smaller than a Gaussian distribution. Again, with a multiplicative noise, the linear stochastic model is able to recover such a sub-Gaussian PDF (Panel (i)). The multiplicative noise coefficient is shown in Panel (h), which clearly illustrates a state dependency.

	\section{Setups in the data assimilation}\label{Sec:DA_Setup}
	
	\subsection{Basic hyperparameters in the data assimilation}
	The setups in the data assimilation are as follows. In the standard test, the ensemble size is $200$, but the data assimilation skill as a function of the ensemble number will also be explored. The localization radius is $38$km for all the state variables. Note that for the localization of the floe position $\mathbf{x}$ in the Lagrangian coordinate, the true observational value is always manually included in the data assimilation \cite{sun2019lagrangian}, since otherwise the data assimilation can become very biased. Although the ocean forecast model is determined by its Fourier coefficients, the solution is transformed to the physical space in each analysis step. Therefore, all the localizations are carried out in the physical space. The observational quantities are the position $\mathbf{x}=(x,y)$ and angular displacement $\Omega$ of each floe. In practice, the error in observing the floe location is about a few hundred  meters while that in observing the angular displacement is around $20^o$. Therefore, Gaussian random noises with standard deviations of $300$ meters and $20^o$ are imposed as the observational error of these two variables, respectively.
	
	\subsection{Simplification of the surface stress integration in the data assimilation forecast step}
	The surface integration of stresses and associated torques are needed in solving the floe model \eqref{eqfirst}. In generating the true signal, a fixed refined  mesh grid with  $250$m$\times250$m resolution is utilized for all the floes to match the resolution of satellite reflectance observations. This means that on average $65\times65$ to $140\times140$ grid boxes are used for each floe, depending on its size. In data assimilation, an adaptive but coarse mesh is adopted for computing the surface integral that saves a significant amount of the computational cost. The number of the grid boxes is fixed as $15\times15$ but the resolution of the boxes changes as a function of the floe size. Such a simplification saves a large amount of the computational cost while remaining the surface integration to be sufficiently accurate for the purpose of data assimilation.
	
	\subsection{Setup of the experiments}
	In the following data assimilation experiments, only observations of non-interacting floes are considered, even though there could be a much larger number of interacting floes present. The number of the non-interacting floes within a $200$km$\times200$km domain in the marginal ice zone of the Beaufort sea that can be detected is about $20$ to $30$ every day \cite{lopez2019ice}. Therefore, in the standard setup here the floe number is $30$. The data assimilation skill with different numbers of the observed non-interacting floes will also be studied since more floes can be identified with the improvement of the satellite resolutions as well as the improved identification methods. Each data assimilation experiment is carried out for 50 days, mimicking the boreal summer, where the skill scores are computed based on the results from day 10 to day 50 to exclude the artificial error during the initial burn-in period. In the standard tests, the observational frequency, i.e., the observational time step, is every 1 day. This is consistent with typical availability of satellite observations \cite{lopez2021library},  although more frequent observational data start to become available. The observational variables are both the linear displacement $\mathbf{x}=(x, y)$ and the angular displacement $\Omega$. One simplification made in the standard setup is that the floes are assumed to have no interactions with each other during their motions. In other words, the collisions between floes are ignored and the floes can `intersect' with each other. The welding and fracturing are also not included here. Therefore, the total number of the floes in the domain equals to the total number of non-interacting floes, which is $30$ throughout the time. Without the collisions, the shape and the thickness of the floes are also assumed to be unchanged. These simplifications facilitate the study of the data assimilation skill as a function of the number of floes by excluding many random effects. In the last part of the experiment section, the elastic collisions will be introduced and the number of the non-interacting floes will be smaller than the total number of the floes in the domain. In this more complicated but realistic situation, the number of non-interacting floes will also vary in time. In addition, each floe will have multiple short periods that have no interactions with others.
	
	\subsection{Skill scores}
	The two skill scores adopted here to assess the data assimilation skill are the pattern correlation (Corr) and the normalized root-mean-square error (RMSE) between the truth (also known as the reference solution) and the assimilated states. For the conciseness of presentation, the RMSE below always stands for the normalized RMSE. The Corr and RMSE are defined as the follows \cite{hyndman2006another, benesty2009pearson},
	\begin{equation}\label{SkillScores}
	\begin{split}
	\mbox{Corr} &= \frac{\sum_{i=1}^I(u^{DA}_{i}-\bar{u}^{DA})(u^{ref}_i-\bar{u}^{ref})}{\sqrt{\sum_{i=1}^I(u^{DA}_{i}-\bar{u}^{DA})^2}\sqrt{\sum_{i=1}^I(u^{ref}_{i}-\bar{u}^{ref})^2}},\\
	\mbox{RMSE} &=  \frac{1}{\mbox{std}(u^{ref})}\left(\sqrt{\frac{\sum_{i=1}^I(u^{DA}_{i}-u^{ref}_{i})^2}{I}}\right),
	\end{split}
	\end{equation}
	where $u^{DA}_{i}$ and $u^{ref}_{i}$ are the posterior mean estimate from data assimilation and the truth of $u$, respectively, at a single point $i$. The value $I$ is the total number of points for computing these skill scores. Depending on the context, $I$ can be the totally number of points in a time series, the total number of the spatial grid points at a fixed time, or the total number of the points in both time and space. The averages of the forecast and the true time series are denoted by $\bar{u}^{DA}$ and $\bar{u}^{ref}$ while std$(u^{ref})$ is the standard deviation of the truth. The truth here can be the solution of a Fourier coefficient or that in the physical domain.
	The RMSE starts from RMSE $=0$ and loses its skill as it increases. The pattern correlation starts from Corr $=1$ and loses its skill as it decreases. The data assimilation results are regarded as skillful if the corresponding RMSE $<1$ and Corr $>0.5$.
	

	\section{Test Results}\label{Sec:Tests}
	This section shows the data assimilation skill in various scenarios. Unless stated otherwise, the linear stochastic models with multiplicative noise are always utilized as the forecast models for the atmospheric and the ocean modes that are highly non-Gaussian. The words `filtered', `recovered' and `assimilated' are interchangeable in the following discussions.
	\subsection{Data assimilation skill in the standard test}\label{Sec:StandardTest}
\begin{figure}
		\hspace*{-0.5cm}\includegraphics[width=18.0cm]{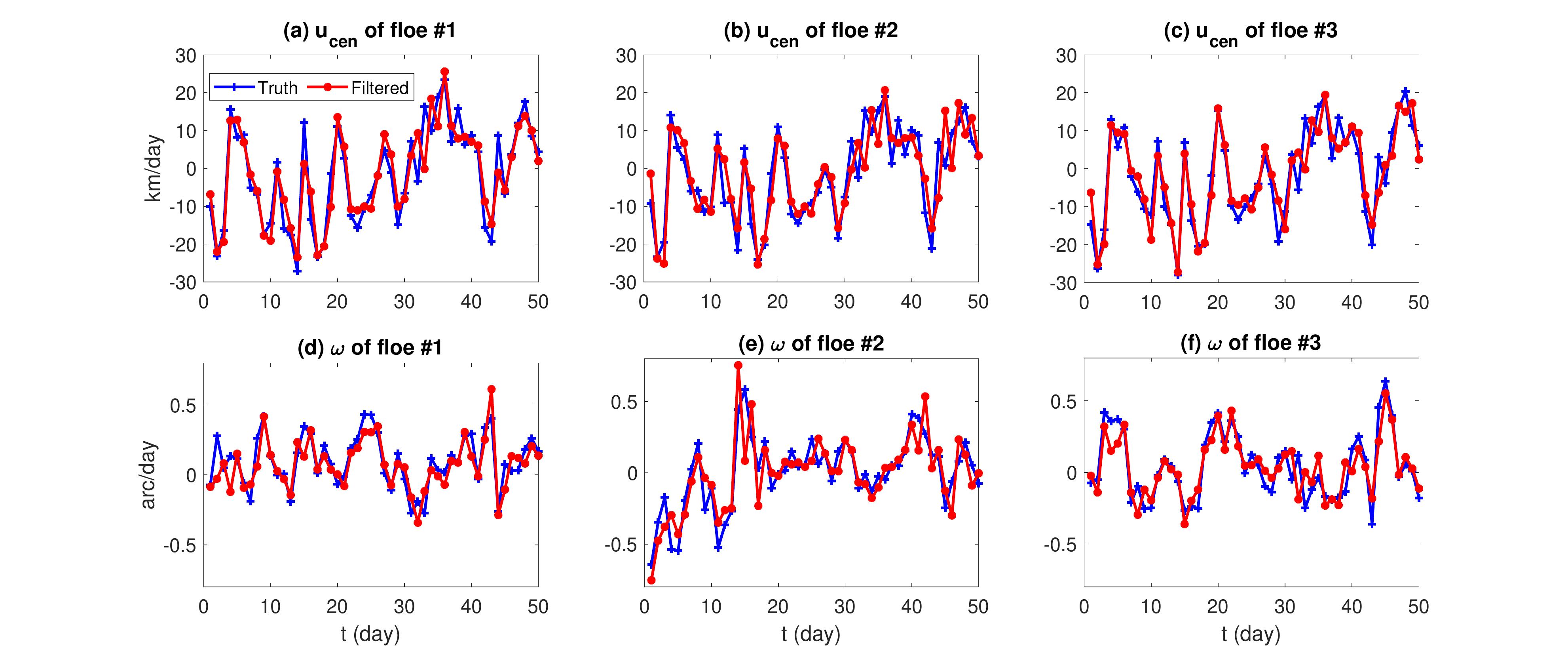}
		\caption{Comparison of the true signals and the filtered ones for the $x$-component of the floe velocity ${\bf v}_{\text{cen}}$ and the angular velocity $\omega$. The three floes here correspond to the three ones labeled in Figure \ref{Flow_Ocean_Illustration}.}\label{Floe_Vel_Recovered}
	\end{figure}

	Recall that in the standard test, $30$ observed floes are utilized, where both the floe positions $\mathbf{x}$ and the angular displacements $\Omega$ are the observed quantities. The observations are available every 24 hours. Figure \ref{Floe_Vel_Recovered} shows the recovered floe velocity ${\bf v}_{\text{cen}}$ in the $x$-direction ($y$-velocity not shown) and the recovered angular velocity $\omega$ for the three floes marked in Figure \ref{Flow_Ocean_Illustration}. The velocities of the three floes evolve similarly in time (Figure \ref{Floe_Vel_Recovered}, a-c), indicating the predominance of the relatively heterogeneous winds in translating the ice. However, the patterns of angular velocity evolution are very different (Figure \ref{Floe_Vel_Recovered}, d-f) because those are dominated by the heterogeneous ocean eddies. Since the floe velocities and angular velocities are directly linked with the observations of their coordinates and angular displacements, the recovered states are quite accurate. The quality of velocity reconstructions for other floes is similar so the associated results are not shown but included in the calculation of the skill metrics.
	
\begin{figure}
		\hspace*{-0.5cm}\includegraphics[width=18.0cm]{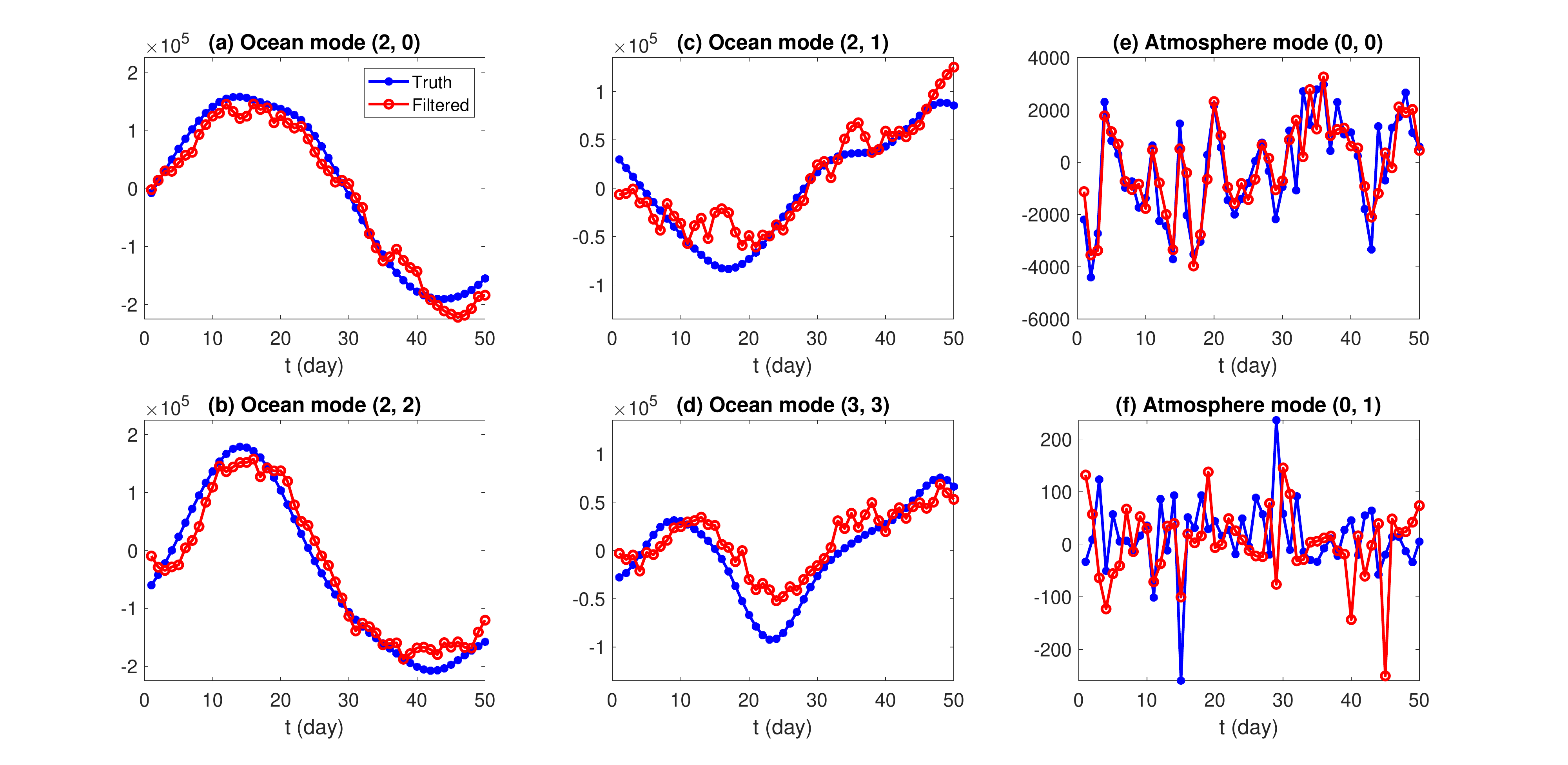}
		\caption{Comparison of the true signals and the filtered ones for the four energetic ocean Fourier modes and the two leading atmosphere modes. Only the real part of the time series is shown here.}\label{Ocean_Fourier_Recovered}
	\end{figure}

	Figure \ref{Ocean_Fourier_Recovered} illustrates the data assimilation skill of recovering the Fourier coefficients associated with the most energetic modes of the ocean and atmosphere flow fields. Clearly, the ocean varies much more slowly than the atmospheric component. The recovered time series of the most energetic ocean modes are quite close to the truth. For the atmosphere leading mode $\mathbf{k}=(0,0)$, the data assimilation via the approximate linear stochastic model with multiplicative noise also gives high recovery skill. For the more intermittent and higher-frequency atmospheric mode $\mathbf{k}=(0,1)$, the recovered time series is not perfect but it roughly captures the tendency and the overall amplitude of the truth. In Section \ref{Sec:Comparison_G_NG}, a comparison between using the linear stochastic models with additive and multiplicative noises will be carried out.

\begin{figure}
		\hspace*{-0.5cm}\includegraphics[width=18.0cm]{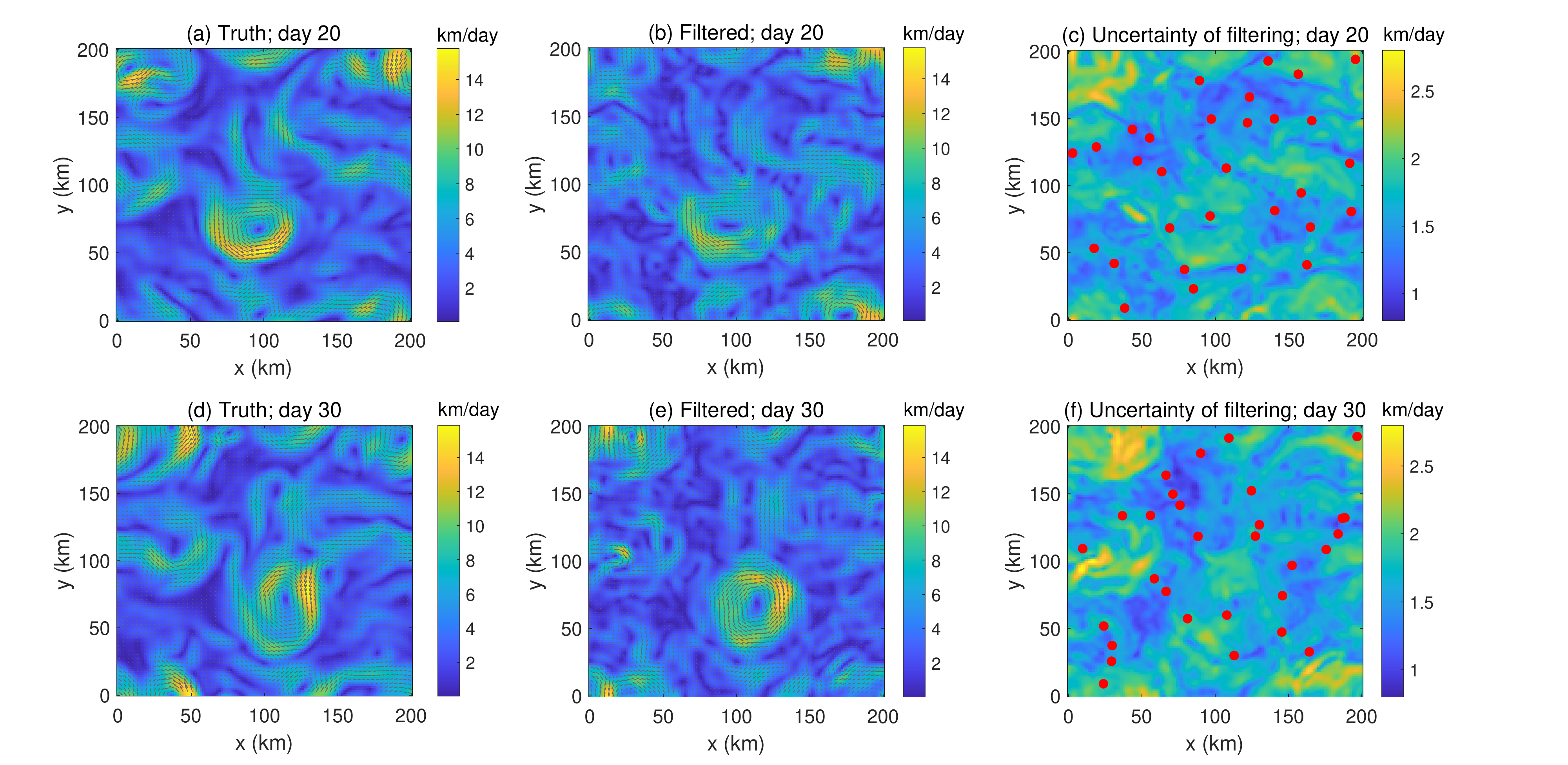}
		\caption{Comparison of the true ocean velocity fields in physical space and the filtered ones at day 20 (Panels (a)--(b)) and day 30 (Panels (d)--(e)). The arrows are the ocean upper layer velocity field while the shading area is the amplitude $\sqrt{u^2+v^2}$ of the velocity. The shading area in Panels (c) and (f) shows the filtering uncertainty, in the form of the square root of the posterior variance at each ocean grid point. The red dots are the mass center of the floes. }\label{Ocean_Truth_Recovered}
	\end{figure}

	The comparison of the reconstructed ocean velocity fields from the filtered solution with the truth is shown in Figure \ref{Ocean_Truth_Recovered}. At both day 20 and day 30, the reconstructed ocean field captures the main structures of the truth. In particular, the large-scale vortex in the center of the domain is clearly recovered. In addition, the regions with strong or weak signals in the recovered flow field are consistent with the truth. In the third column of this figure, the uncertainty associated with the data assimilation  is included. The  red dots show the mass center of the floes. The amplitude of the uncertainty is on average less than $2$(km/day), which is much smaller than the strength of the velocity field that is $10$ to $14$km/day. This means the recovered velocity field from the data assimilation is trustable. It is also worthwhile to point out that the floes are nearly uniformly distributed in the domain. This is consistent with the previous theoretic study when the underly ocean field is incompressible \cite{chen2015noisy}. Finally, the uncertainty of the recovered ocean field is overall larger at the locations with no observed floes nearby than the uncertainty in the areas that are surrounded by the sea ice floes. This is due to the application of the localization in data assimilation. In fact, if there is no observed floes within the localization radius, then the data assimilation simply trusts the model forecast results, which is overall less accurate than the combination of model and data. Note that, if there is no localization and the ensemble size is large enough (which is however computationally expensive), then it is expected that the uncertainty should be statistically homogeneous in the domain since the ocean field at different locations is globally coupled.

	The results in this subsection indicates that $30$ observed floes are sufficient to provide a reasonable recovery of the turbulent ocean flow field.

	\subsection{Data assimilation skill with different parameter setups}
\begin{figure}
		\hspace*{-0.5cm}\includegraphics[width=18.0cm]{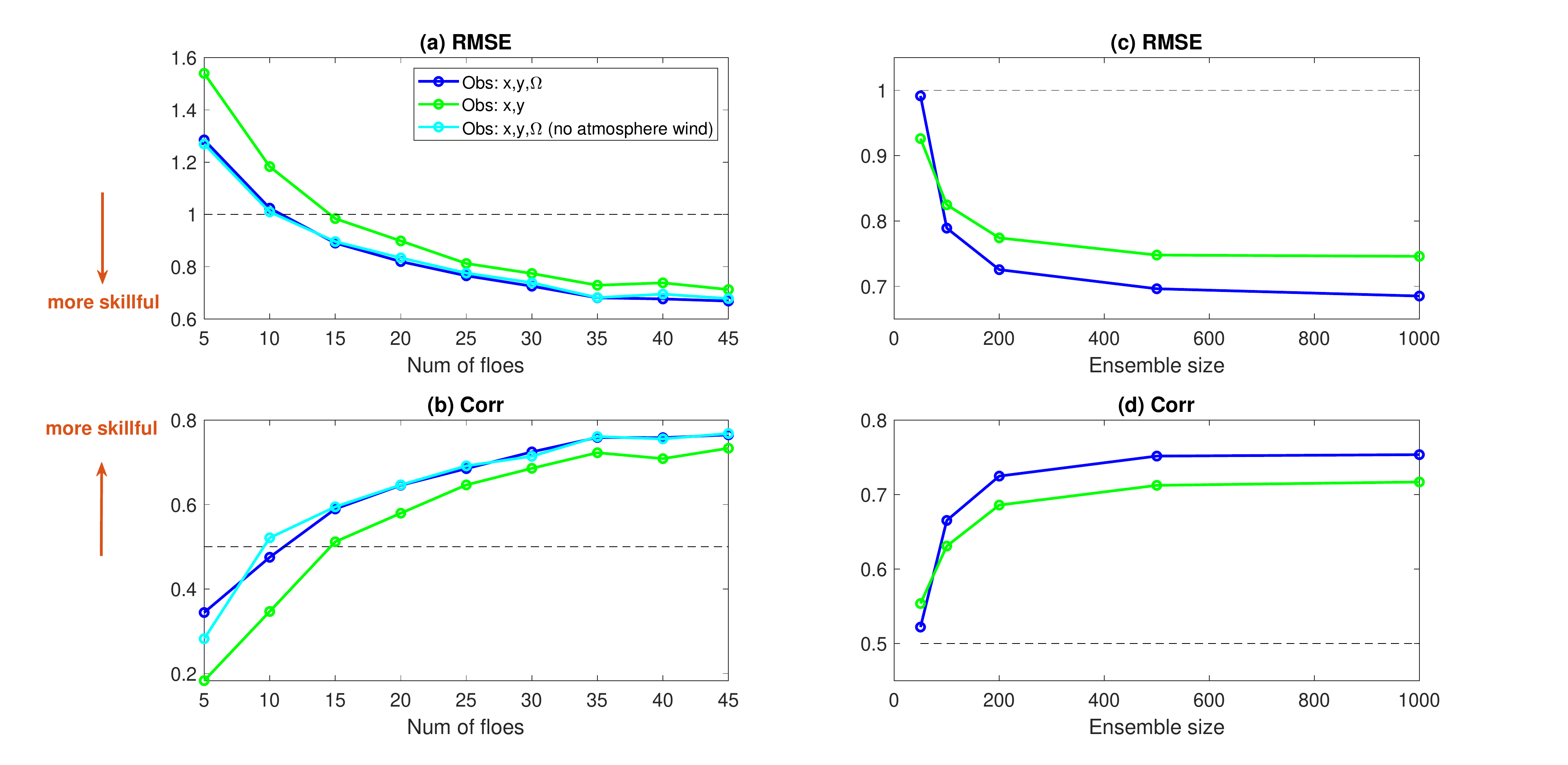}
		\caption{Panels (a)--(b): Skill scores as a function of the number of floes, where the number of the ensemble is fixed $=200$. Panels (c)--(d): Skill scores as a function of the ensemble size, where the number of floes is fixed $=30$. The skill scores here are computed between the true ocean flow field and the recovered one transformed back to the physical space. Since the velocity field is a two-dimensional vector, the skill scores are computed based on the amplitude of the flow field $\sqrt{u^2+v^2}$ averaged over both time and space. The skill scores for the two velocity components have similar behavior. }\label{Scores_FloesNum_and_EnsSize}
	\end{figure}

	Figure \ref{Scores_FloesNum_and_EnsSize} shows the data assimilation skill scores as a function of different parameter setups. The skill scores here are computed between the true ocean flow field and the recovered one transformed back to the physical space. Since the velocity field is a two-dimensional vector, the skill scores are computed based on the amplitude of the flow field $\sqrt{u^2+v^2}$ averaged over both time and space. In the left column, the skill scores as a function of the number of the observed floes are illustrated. If $x$, $y$ and $\Omega$ are all observed, then the data assimilation is skillful as long as more than $10$ are observed. The improvement of the data assimilation skill from using $25$ observed floes to $45$ is not very significant, indicating that the current observational network ($20$ to $30$ non-interacting floes) is at the turning point in providing reasonable data assimilation skill.
	
	Recall that one significant feature of the sea ice floe observations compared with the traditional Lagrangian tracers is that the floes can provide additional observational information from the angular displacement. Comparing the green and the blue curves in the left column indicates that the observed angular displacement indeed benefits the data assimilation. In fact, the roles of the angular displacement and the positions are similar in reducing the error and the uncertainty in data assimilation. For example, the data assimilation using $10, 20$ and $30$ floes by observing $(x,y,\Omega)$ is comparable to that using $15, 30$ and $45$ floes by observing only $(x,y)$, respectively. On the other hand, if the atmospheric wind is excluded in the system, then the data assimilation skill (cyan curves) for the ocean remains almost unchanged. This is because the atmospheric wind imposed here is only at the large scale, which does not interact too much with the internal variability of the ocean.
	
	The right column of Figure \ref{Scores_FloesNum_and_EnsSize} shows the data assimilation skill as the number of the ensemble size, which indicates that $200$ ensembles is a suitable choice for the LETKF here considering the computational cost and accuracy. A further decrease of the ensemble size will deteriorate the data assimilation skill.

	\subsection{Data assimilation skill for recovering the ocean flow field at different spatial scales}
\begin{figure}
		\hspace*{-0.5cm}\includegraphics[width=18.0cm]{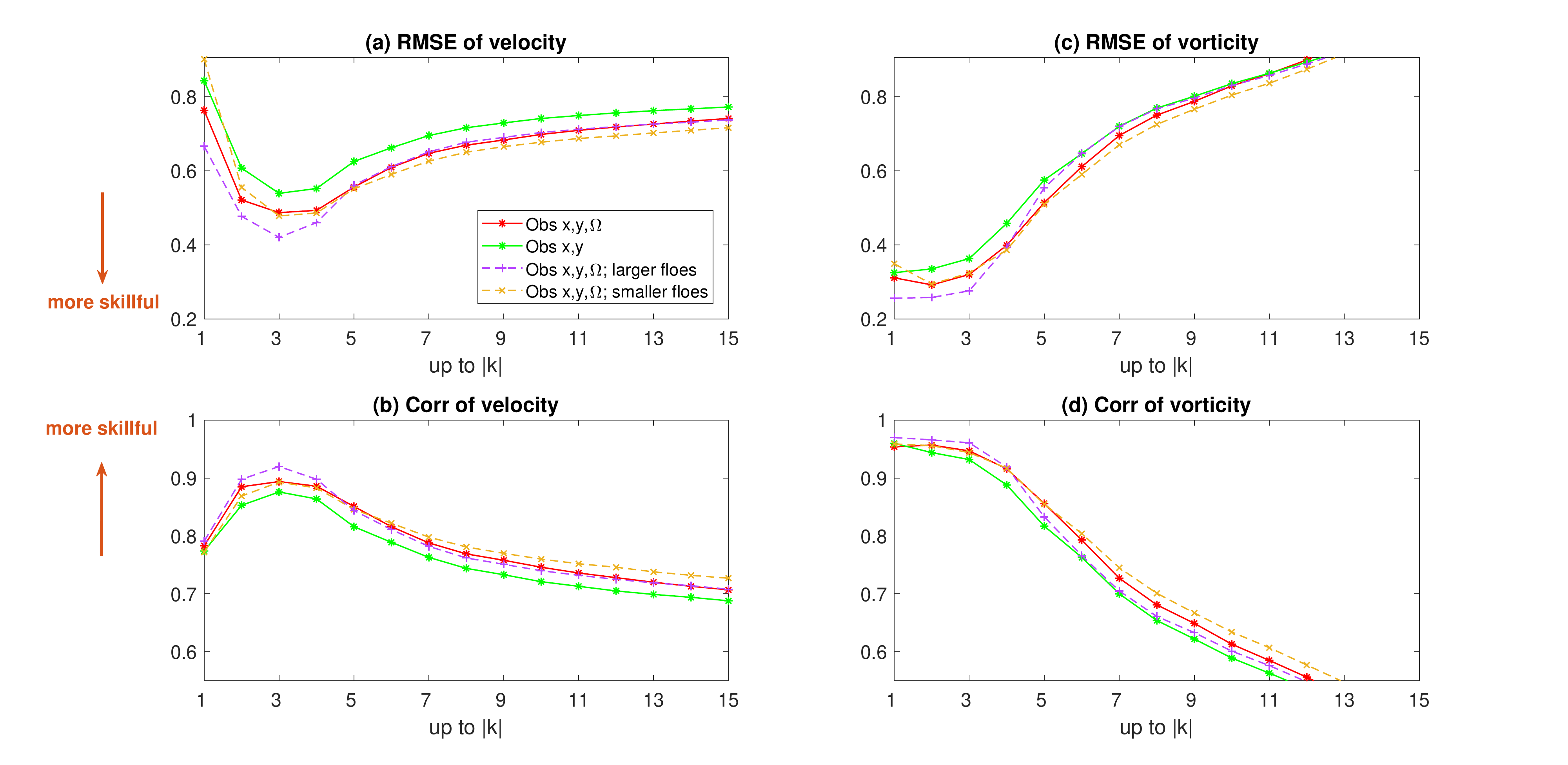}
		\caption{Panels (a)--(b): Skill scores of the ocean velocity field as a function of the number of the spatial scale included in the reconstruction. Panels (c)--(d): Skill scores of the ocean vorticity field as a function of the number of the spatial scale included in the reconstruction. The solid red and green curves show the situation with the floe size the same as in the standard setting. The dashed magenta and yellow curves show the data assimilation with $30$ larger and $30$ smaller floes as observations, respectively. The larger floes have radii that range roughly from $21$km to $27$km while the smaller floes have radii that range roughly from $4$km to $7$km.}\label{Skill_Scores_Vel_Vort}
	\end{figure}

	So far, the study focused on the overall data assimilation skill for recovering the entire ocean field. In this subsection, the data assimilation skill at different spatial scale is studied. Figure \ref{Skill_Scores_Vel_Vort} shows the skill scores as a function of the spatial scale included in the reconstructed flow field in the physical domain using Fourier wave numbers up to $|\mathbf{k}|$. Regardless of the difference in choosing the observations in the experiments studied here, the most skillful range of the recovered flows is at the spatial scales from $|\mathbf{k}|=2$ to $|\mathbf{k}|=5$. Note that radii of most of the floes are between $8$km to $18$km while the entire domain is $200$km$\times200$km. These facts confirm the finding that the floe size should be smaller than the targeted spatial scale since otherwise the detailed information of the underlying flow is averaged out by the surface integration. In fact, if a set of larger floes are used (with radii ranging roughly from $21$km to $27$km), then the data assimilation skill of the large scales will be improved. In contrast, if set of larger floes are used (with radii ranging roughly from $4$km to $7$km), then the results of the small-scale ocean features will be recovered more accurately. On the other hand, the data assimilation using only $(x,y)$ as the observations is less skillful compared with the case by observing $(x,y,\Omega)$, which is consistent with the results shown in Figure \ref{Scores_FloesNum_and_EnsSize}. Finally, the ocean velocity is recovered more accurately than the vorticity. This is because vorticity requires one more derivative compared with the velocity and the error in the small scales are amplified in the spatial reconstruction.
	
	\subsection{Data assimilation with missing observations due to cloud cover}
	One potential difficulty in the Lagrangian data assimilation of sea ice is the presence of the cloud cover that obscures floe observations (Figure \ref{Satellite_Image}). These cloud covers can sometimes be as big as the size of half or the entire domain and can also last for  several  days. To study the data assimilation in the presence of cloud covers, define $L_0$ as the number of floes (out of $L$ total) that are covered by clouds. Assume a big chunk of the cloud appears which covers a large part of the domain. These $L_0$ unobserved floes are thus clustered in a certain area of the domain instead of randomly distributed. Since the large-scale motion of the floes are mostly driven by the atmospheric wind, the cloud is assumed to move together with the floes. In other words, the same $L_0$ floes are unobserved during the period that the cloud covers appear.
	
\begin{figure}
		\hspace*{-0.5cm}\includegraphics[width=18.0cm]{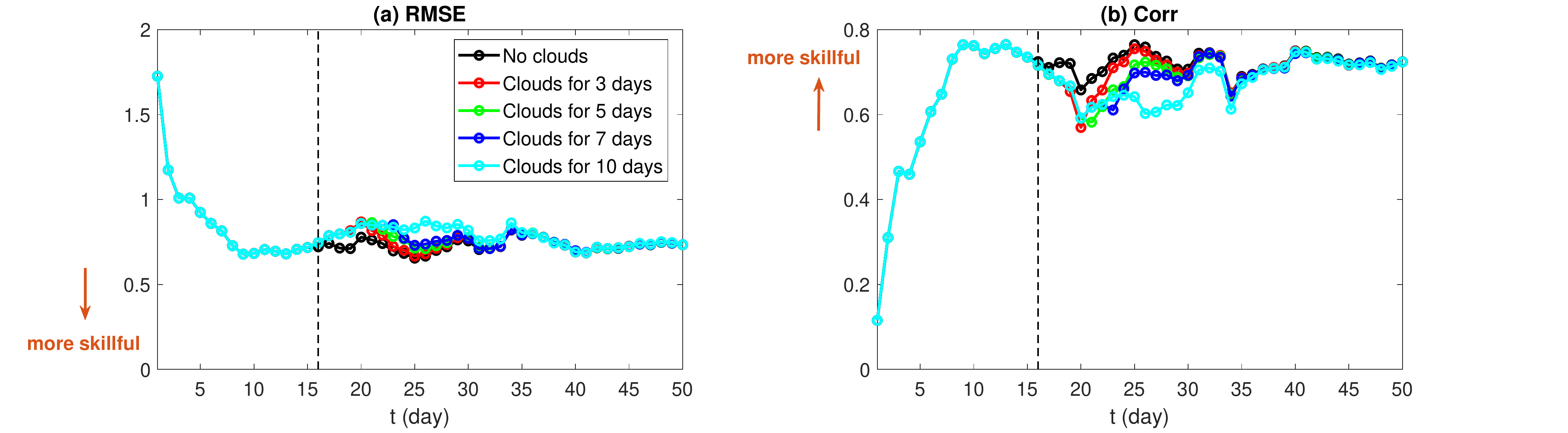}
		\caption{Skill scores as a function of time. Starting from day 16, a large piece of clouds appear. The curves with different colors show the skill scores in which the clouds stay for 3 (red), 5 (green), 7 (blue) and 10 (cyan) days. The black curves show the case that there is no cloud throughout the entire period. }\label{Skill_Scores_Clouds}
	\end{figure}

	In the following experiment, the total number of the floes that are observed without the cloud is still $L=30$. Starting from day 16, $L_0=15$ floes are obscured due to the cloud covers, which occupies half of the entire domain. Figure \ref{Skill_Scores_Clouds} shows the skill scores as a function of time (days), where the curves in different colors correspond to the situations for different length of days with cloud covers. With the appearance of the clouds, the number of observations starting from day 16 decreases to only $15$, and therefore the data assimilation skill becomes worse. As is expected, the longer the clouds last, the less skillful the recovered state will be after day 16. The results in this figure also show that it will take several days for the data assimilation skill to adjust back to the situation as if there is no cloud cover throughout the period.
	
\begin{figure}
		\hspace*{-0.5cm}\includegraphics[width=18.0cm]{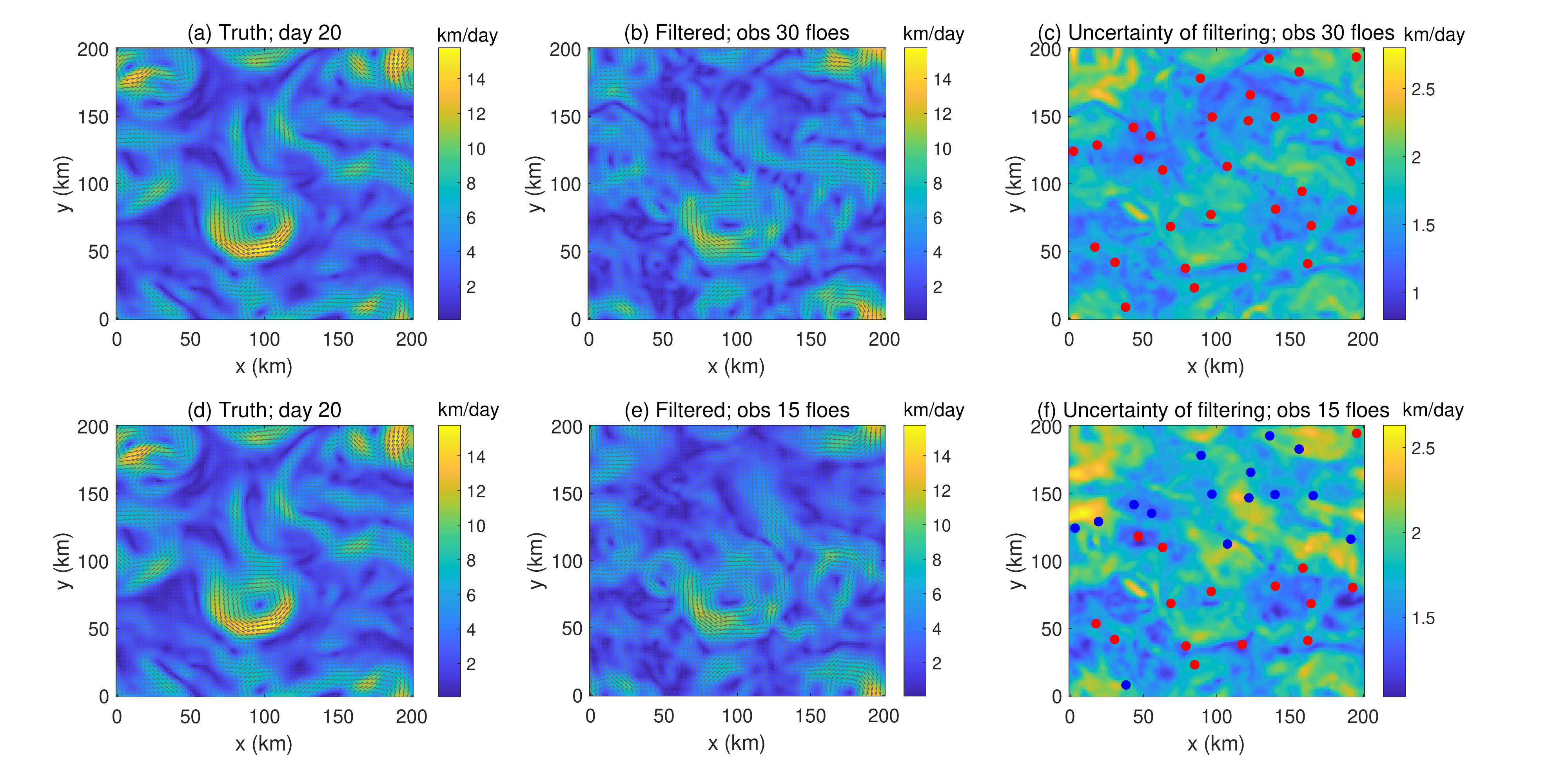}
		\caption{The top row shows the situation with $30$ observed floes throughout the period. The bottom row illustrates the case that $15$ floes are unavailable due to the clouds starting from day 16, where the clouds last for 10 days (corresponding to the cyan curves in Figure \ref{Skill_Scores_Clouds}). The results at day 20 are shown. 
			In Panels (a)--(b) and (d)--(e), the arrows are the ocean upper layer velocity field while the shading area is the amplitude $\sqrt{u^2+v^2}$ of the velocity. The shading area in Panels (c) and (f) shows the filtering uncertainty, in the form of the square root of the posterior variance at each ocean grid point. The red dots are the mass center of the floes. The blue dots are the unobserved floes covered by clouds while the red ones are the observed ones.}\label{Filter_with_Cloud}
	\end{figure}

	Figure \ref{Filter_with_Cloud} compares the truth and the recovered ocean field at day 20. The top row shows the situation with $30$ observed floes throughout the period. The bottom row illustrates the case that $15$ floes are unavailable due to the clouds starting from day 16, where the clouds last for 10 days (corresponding to the cyan curves in Figure \ref{Skill_Scores_Clouds}). Panels (c) and (f) of Figure \ref{Filter_with_Cloud} include the uncertainty associated with the recovered ocean field together with the locations of the floes, where the blue dots are the unobserved floes covered by clouds while the red dots are the observed ones. Comparing Panel (e) with Panel (b) of Figure \ref{Filter_with_Cloud}, it is clear that the recovered flow field in Panel (e) is less accurate, which is consistent with the skill scores shown in Figure \ref{Skill_Scores_Clouds}. The error grows rapidly in the upper half of the domain, which is covered by the clouds. Likewise, the overall uncertainty in Panel (f) is bigger than that in Panel (c). The most significant increment of the uncertainty in day 20 again occurs in the upper half of the domain, where there is no observed floes. The reason of such a distinguished behavior in the recovered flow field in the upper and lower haves of the domain is the same as those discussed at the end of Section \ref{Sec:StandardTest}, which is due to the application of the localization in data assimilation.

	\section{Discussions}\label{Sec:Discussions}
	\subsection{Model error in data assimilation utilizing the reduced order forecast model}
\begin{figure}
		\hspace*{-0.5cm}\includegraphics[width=18.0cm]{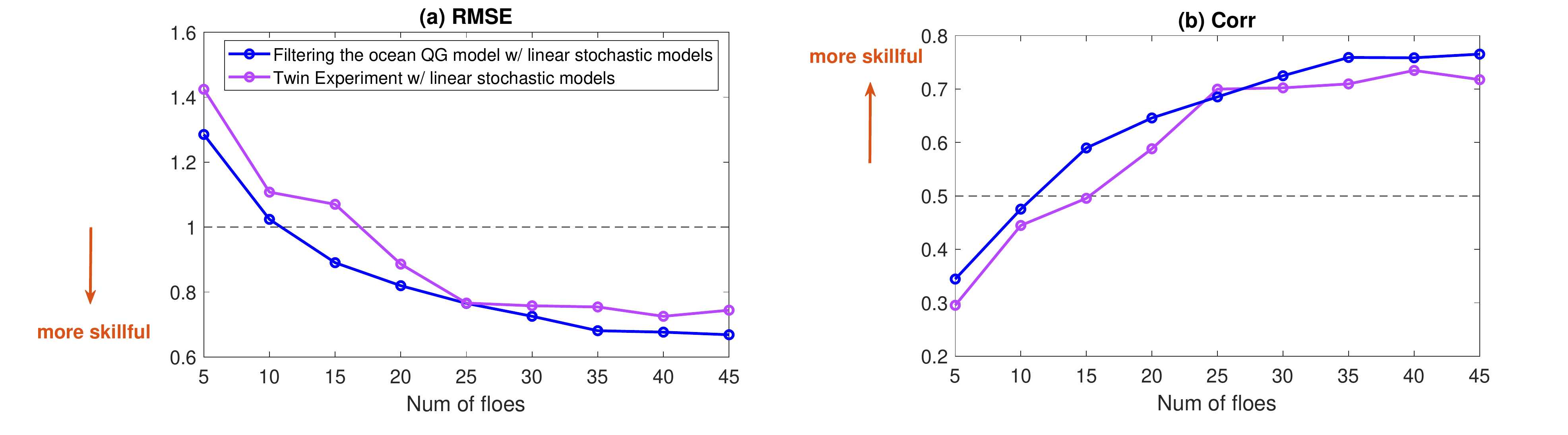}
		\caption{Comparison of the skill scores as a function of the number of floes in two experiments. The blue curves show the results corresponding to setup in the previous section, where the QG model is used to generate the true signal while the linear stochastic models are used as the forecast model in data assimilation. The magenta curves show the results based on a twin experiment, where the linear stochastic models are used to both generate the true signal and serve as the forecast model in data assimilation.}\label{Model_Error_Check}
	\end{figure}

	One natural topic to explore in using the set of linear stochastic equations as the forecast model is the model error in data assimilation. However, using the complete QG ocean model as the forecast model for data assimilation is computationally unaffordable in the sense of both the computational time and the computational storage. This prevents running a perfect twin experiment for understanding the model error. The study of the model error here is thus based on a twin experiment in a slightly different way, where the linear stochastic models are used to both generate the true signal and serve as the forecast model in data assimilation. The resulting data assimilation skill scores are overall similar and slightly worse than the one in which the QG ocean model is used to generate the true signal while the linear stochastic models are adopted as the forecast models. See Figure \ref{Model_Error_Check}. In fact, the true signal generated from a simple stochastic model with the same level of the uncertainty as the nonlinear deterministic model is in general slightly harder to be predicted and assimilated due to the lack of a clear dynamics. Nevertheless, the comparable data assimilation skill scores here at least indicate that the linear stochastic models are good approximations for the QG model in data assimilation.
	
	\subsection{Comparison of the linear forecast models with additive and multiplicative noise}\label{Sec:Comparison_G_NG}
	In Figure \ref{Time_Series_Fits}, it has been shown that the linear model with multiplicative noise is crucial in capturing the non-Gaussian behavior of the atmospheric modes. In this subsection, the role of the multiplicative noise for data assimilation is studied.
	
\begin{figure}
		\hspace*{-0.5cm}\includegraphics[width=18.0cm]{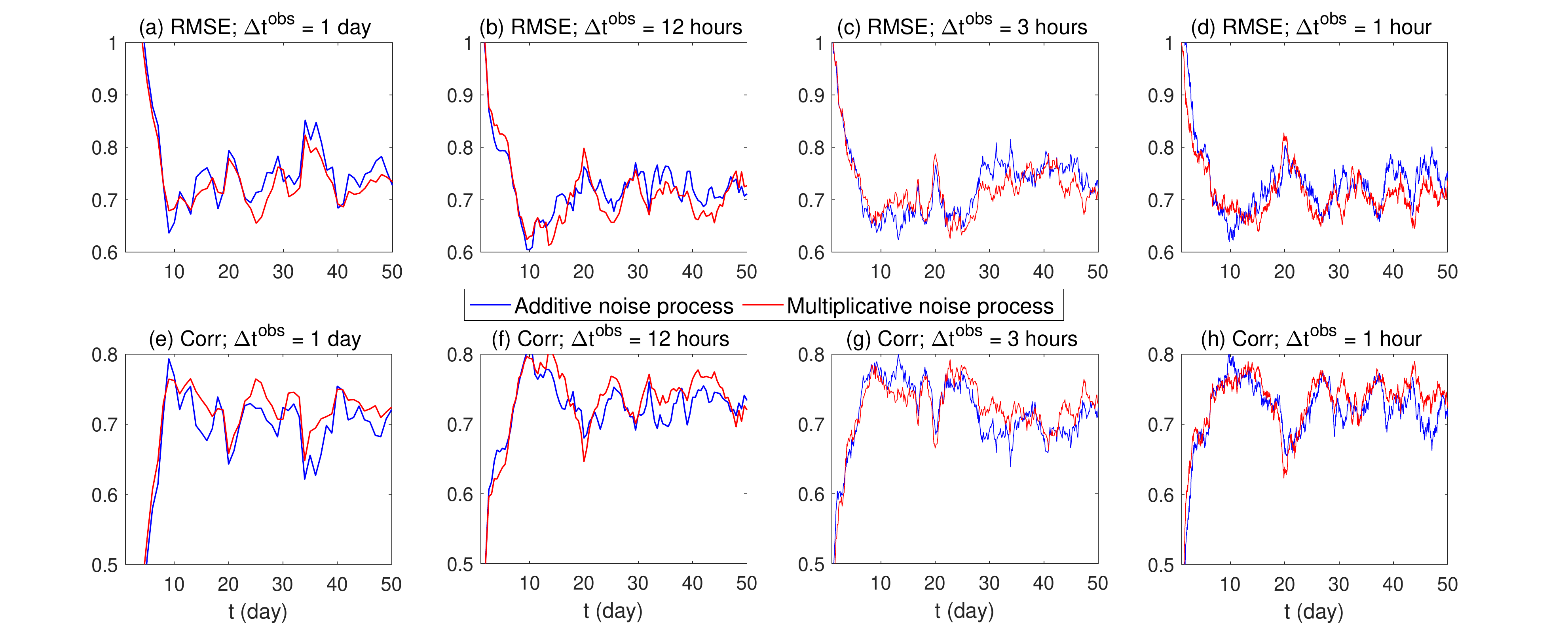}
		\caption{The skill scores of the recovered ocean field as a function of time (day). Different columns show the experiments with different observational time steps for one day, 12 hours, 3 hours and 1 hour, respectively. The blue curves show the experiments, where the linear models with additive noise are used for all the atmospheric and ocean modes. In contrast, the red curves show the cases, where the linear models with multiplicative noise are used for the atmospheric modes. In both experiments, all the ocean modes are all modeled by linear models with additive noise.  }\label{Compare_G_nG_SkillScore}
	\end{figure}

	Figure \ref{Compare_G_nG_SkillScore} compares the skill scores of the recovered ocean field utilizing different approximate forecast models. The blue curves show the experiments, where the linear models with additive noise are used for all the atmospheric and ocean modes. In contrast, the red curves show the cases, where the linear models with multiplicative noise are used for the atmospheric modes. In both experiments, all the ocean modes are modeled by linear models with additive noise. Different columns in Figure \ref{Compare_G_nG_SkillScore} show the experiments with different observational time steps, for one day, 12 hours, 3 hours and 1 hour, respectively. Despite the fluctuations in the time evolution of the skill scores due to the random and turbulent effects, the recovered ocean fields using the linear stochastic model with multiplicative noise as the forecast model are overall more accurate than its counterpart that involves only an additive noise process, regardless of the observational frequency. The reason of the more skillful data assimilation results utilizing the model with multiplicative noise is that it can better forecast the uncertainty at the transient phases induced by the intermittencies, which is often underestimated by linear models with additive noise.
	As a result, the update step in the data assimilation using the multiplicative noise model is more skillful in combining the information from the forecast model and that from the observations. Such a role of the  multiplicative noise  has also been explored in the stochastic parameterization of climate models \cite{berner2017stochastic, sura2005multiplicative}. Note that noise (or covariance) inflation techniques \cite{hamill2005accounting, anderson2009spatially} are widely utilized in practice with successes. An increased noise coefficient aims at compensating the underestimation of the uncertainties in the linear models with additive noise, which plays a similar role as the multiplicative noise process utilized here. However, the noise inflation often requires many empirical tunings while the procedure of determining the multiplicative noise coefficient here is systematic.
	
\begin{figure}
		\hspace*{-0.5cm}\includegraphics[width=18.0cm]{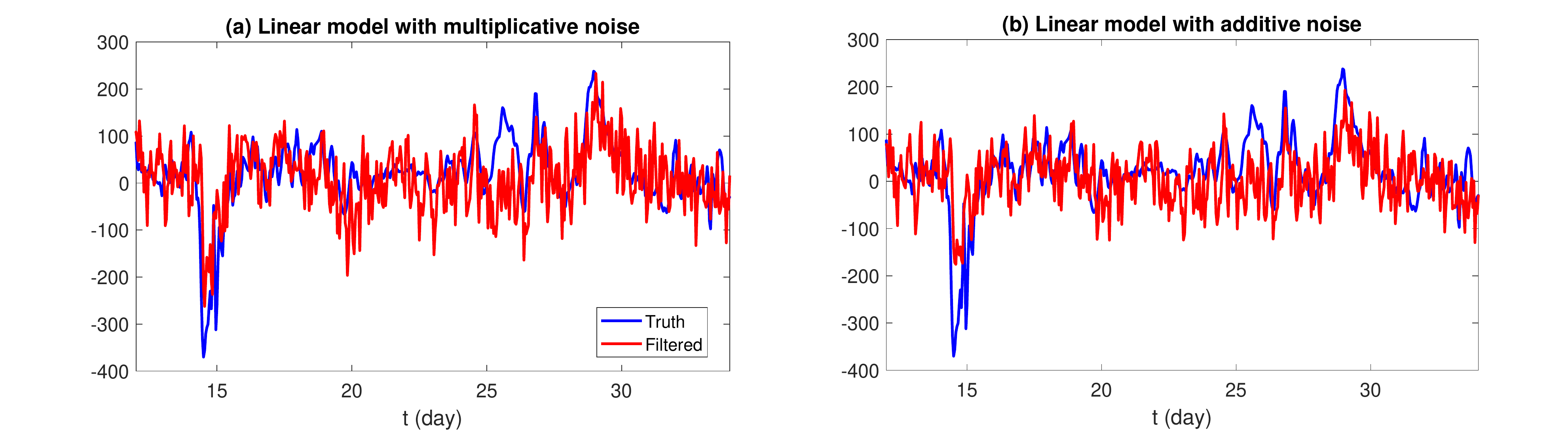}
		\caption{Comparison of the data assimilation skill of the atmosphere mode $(0,1)$ of the zonal velocity $u$ using the linear model with multiplicative noise (Panel (a)) and that using the linear model with additive noise (Panel (b)). Here the observational time step is every $1$ hour. Only the real part of the time series is shown.}\label{Compare_G_nG}
	\end{figure}

	Figure \ref{Compare_G_nG} compares the data assimilation skill of the atmosphere mode $(0,1)$ associated with the zonal velocity $u$ using the linear model with multiplicative noise (Panel (a)) and that using the linear model with additive noise (Panel (b)). Despite having similar behavior in recovering the quiescent phases, the recovered time series using the linear model with multiplicative noise is much more skillful in recovering the extreme events. Note that the experiment here uses a short observational time step. The difference between these two models becomes less significant if daily observations are used due to the fact that the mode $(0,1)$ has a short decorrelation time (see Panel (d) of  Figure \ref{Time_Series_Fits}). Yet, other modes, for example $(0,0)$, has longer decorrelation times. In fact, the results in Figure \ref{Compare_G_nG_SkillScore} already illustrate the advantage of using the linear model with multiplicative noise for various observational time steps.

	\subsection{The advantage of utilizing linear model with multiplicative noise over a family of nonlinear models in Lagrangian data assimilation}
	Since the most crucial role of the reduced order model is to capture the forecast uncertainty, there are multiple ways of developing approximate reduced order models for data assimilation. In fact, there is a rich family of nonlinear forecast models, which are named as the stochastic parameterized extended Kalman filter (SPEKF) \cite{gershgorin2010improving, gershgorin2010test}, that have been shown to be skillful in many applications \cite{branicki2018accuracy, majda2018model}. A simple version of the SPEKF model is as follows,
	\begin{equation}\label{SPEKF}
	\begin{split}
	\frac{\d \hat{u}_{\mathbf{k}}}{\d t} &= (\gamma_{\mathbf{k}} + i\omega_{\mathbf{k}})\hat{u}_{\mathbf{k}} + \sigma_{\mathbf{k},u}\dot{W}_{\mathbf{k},u}\\
	\frac{\d \gamma_{\mathbf{k}}}{\d t} &= -d_{\mathbf{k},\gamma} (\gamma_{\mathbf{k}} - \hat{\gamma}_{\mathbf{k}}) +  \sigma_{\mathbf{k},\gamma}\dot{W}_{\mathbf{k},\gamma},
	\end{split}
	\end{equation}
	where $\hat{u}_{\mathbf{k}}$ is one of the Fourier modes. The second equation of $\gamma_{\mathbf{k}}$ is a stochastic process that parameterizes the damping of $\hat{u}_{\mathbf{k}}$, which allows the distribution of $\hat{u}_{\mathbf{k}}$ to be non-Gaussian. The stochastic damping $\gamma_{\mathbf{k}}$ is then acted as an augmented state variable, which is updated together with the other physical variables in data assimilation.  The SPEKF has the advantages of utilizing closed analytic formulae for describing the time evolution of the moments. This means if the observations are linear on $\hat{u}_{\mathbf{k}}$, then the entire data assimilation can be solved exactly and accurately. This advantage facilitates the data assimilation with Eulerian observations. However, in the presence of Lagrangian observations, such a merit no long exists. In addition, the observational information needs to pass from the floe displacements to the ocean velocity via the floe locations before arriving at the stochastic process $\gamma_{\mathbf{k}}$. The turbulent nature of the coupled system can lead to issues of the observability and results in an inaccurate estimation of $\gamma_{\mathbf{k}}$. Finally, determining the parameters $d_{\mathbf{k},\gamma}$, $\hat{\gamma}_{\mathbf{k}}$ and $\sigma_{\mathbf{k},\gamma}$ in the SPEKF often requires ad hoc tunings. Each dynamics of Fourier coefficient needs to be described by one individual SPEKF model. Therefore, a systematic model calibration can be a difficult task. In contrast, the new strategy requires a minimum number of parameters in the linear stochastic approximate model and the calibration is systematic.

	\subsection{Data assimilation using the floe model with collisions}
	This  subsection studies the data assimilation skill scores in the presence of elastic floe collisions, which is a more realistic setup for low-concentrated marginal ice zones. This means the observations of each floe are only available in a few disjoint time intervals, during which no collision happens for that specific floe.
	
\begin{figure}
		\hspace*{-0.5cm}\includegraphics[width=18.0cm]{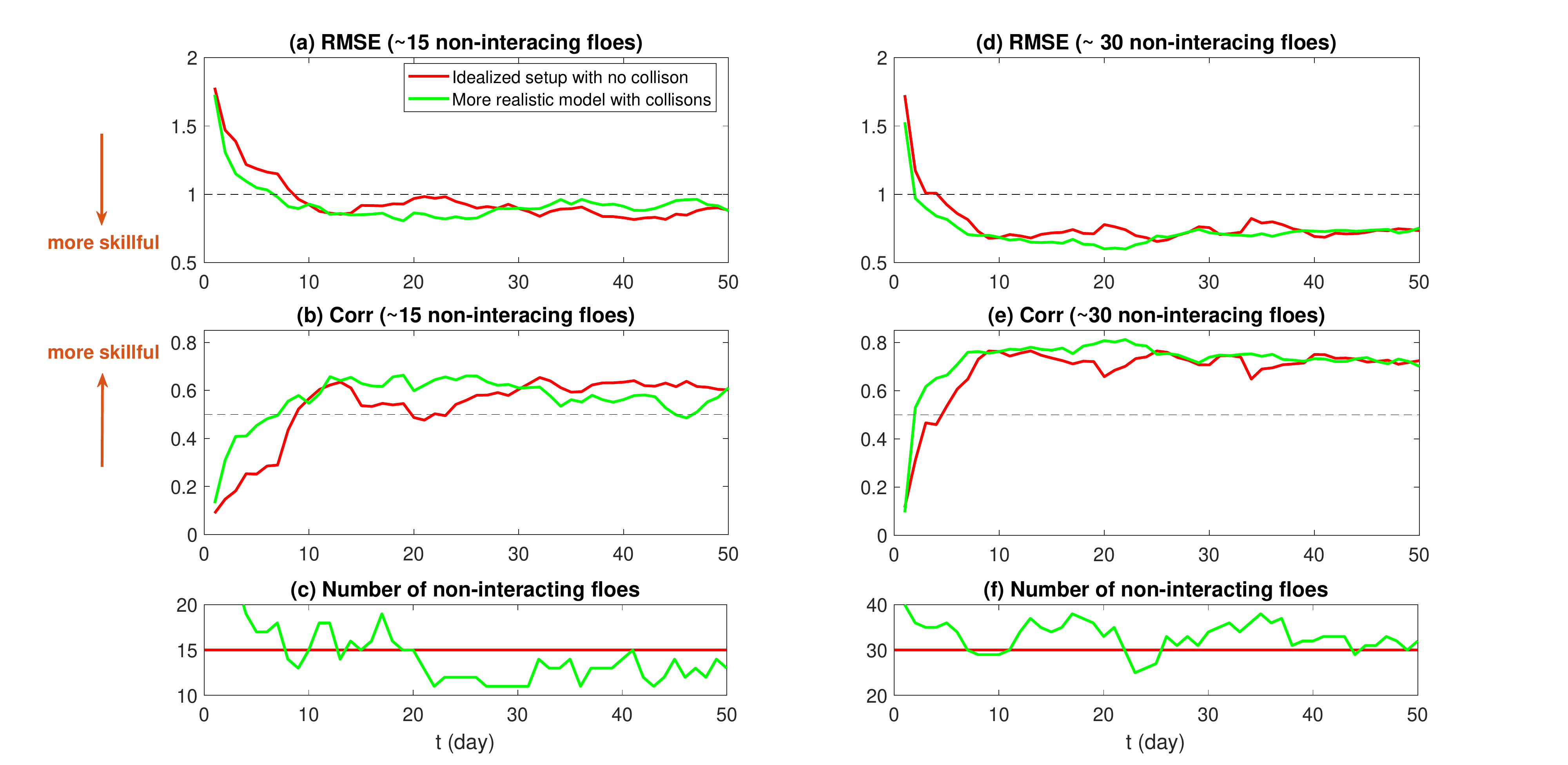}
		\caption{Comparison of the data assimilation skill scores using the idealized setup with no collision (red) and the more realistic model including the collisions (green). A fixed number of $15$ (Panels (a)--(b)) and $30$ (Panels (d)--(e)) floes are utilized in the idealized setup for the cases in the left and the right columns, respectively. In the more realistic setup, $30$ and $50$ floes are included in the domain, which allow the number of the non-interacting floes to be around $15$ and $30$, although the exact number (Panel (c) and Panel (f)) fluctuates in time. }\label{Comparison_Collision}
	\end{figure}

	Figure \ref{Comparison_Collision} compares the data assimilation skill scores based on the idealized setup with no collision as in the previous subsections (red) and the more realistic model including the collisions (green). A fixed number of $15$ (Panels (a)--(b)) and $30$ (Panels (d)--(e)) floes are utilized in the idealized setup for the cases in the left and the right columns, respectively. In the more realistic setup, $30$ and $50$ floes are included in the domain, which allow the number of the non-interacting floes to be around $15$ and $30$, although the exact number (Panel (c) and Panel (f)) fluctuates in time. Note that in the second case (with in total $50$ floes), the floe size is reduced in order to guarantee the number of the non-interacting floes to be around $30$. On average, the numbers of the days that one floe does not interact with others are 24 and 33 (out of 50) days in the situations with $30$ and $50$ floes, respectively.  Here, the day before and the day after the collision are defined as the days between which the floes interact with each other. Note that separating the observations into before and after collisions is only plausible in relatively low-concentrated areas when collisions are rare. Physically, the effect of the collision force on the floe velocity and angular velocity is damped quickly by the ocean and atmospheric drag, within only a couple of hours. Therefore, the collision effect will not play any major role for the data assimilation as long as the observations are splitted into two individual periods with a broken point at the collision instant. It is shown in Figure \ref{Comparison_Collision} that the skill scores are similar to each other in the two model setups when the effective number of the non-interacting floes are the same. Utilizing the non-interacting floes facilitates the data assimilation since the forecast model involving the collision can be much more complicated and computationally expensive. Quantifying the uncertainties resulting from the collision can also be quite challenging.

	\subsection{Recovering the ocean velocity field in the lower layer}
	The reduced order model for the ocean in this study focuses only on the upper layer, since it is the component that couples directly with the sea ice floes. It is nevertheless possible to recover the lower layer of the ocean with a minimum adjustment of the algorithm. One natural idea is to build another set of linear stochastic models that describes the lower layer velocity field of the ocean. The crucial step here is to include the correct statistical coupling between the upper and lower layers. To this end, part of the stochastic noise sources in each linear stochastic model for the lower layer is set to be the same as that in the upper layer. The associated coefficient can be calibrated by the correlation coefficient between the two time series from the original QG model. In such a way, the statistical forecast of the reduced order model reproduces the correlation in the original model. Such a correlation allows the observational information to be passed from the top to the lower layer, which helps improve the state estimation of velocity field of the latter. Alternatively, the lower layer velocity field can be estimated in an offline fashion by building a simple regression model between the velocity fields of the two layers, as was suggested by \cite{molcard2005lagrangian}. Then the data assimilation results of the upper layer can be utilized as the input to recover the flow field of the lower layer.

	\section{Conclusion}\label{Sec:Conclusion}
	In this article, an efficient and statistically accurate Lagrangian data assimilation algorithm is developed. It is then applied to a sea ice model forced by the atmospheric winds and eddying oceanic currents. In this system, the DEM models are utilized to describe the sea ice and the Lagrangian observations are the sea ice floes.
	
	The new data assimilation algorithm contains three main steps (see Figure \ref{Illustration}). First, the oceanic and atmospheric state variables are transferred to the Fourier domain. Second, a reduced order system is developed, which involves only a small portion of the Fourier modes corresponding to the energetic ones. Third, a set of decoupled linear stochastic models are developed to characterize the uncertainty propagation of these modes that are crucial for data assimilation. The stochasticity is utilized to compensate the role of the complicated nonlinearity in creating the uncertainties. The computational cost by running these decoupled stochastic models is significantly cheaper than forecasting the original system. Depending on the long-term statistics of each Fourier coefficient, either an additive noise or a multiplicative noise process is utilized in the associated stochastic model. Analytic formulae are available for determining all the parameters in these linear models, including the multiplicative noise coefficients. Therefore, the entire process of calibrating the reduced order models is systematic and efficient, which avoids empirical tuning.
	
	In the application to the regional sea ice DEM representing a $200$km$\times200$km domain in a marginal ice zone, the data assimilation requires daily observations of at least $10$ non-interacting and uniformly spread out floes to recover the turbulent ocean field with  a Corr$>0.5$. With $30$ non-interacting floes, the ocean field can be recovered with a Corr$>0.7$. In addition to the floe positions, the angular displacements of the floes are very useful to provide extra information that facilitates data assimilation. The results here also indicate that the large/small size of the floes are more skillful in recovering the large/small-scale features of ocean. With the help of the Fourier domain data assimilation, the resulting uncertainty is nearly
	uniformly distributed even in the presence of cloud cover that obscures the observations in a large area. Despite being a much simpler forecast model, it has been shown that the model error in using the linear stochastic models is insignificant in deteriorating the data assimilation skill. Nevertheless, the multiplicative noise in the linear stochastic models is shown to be important in quantifying the forecast uncertainty and recovering extreme events. It has also been shown that the collision effect will not play any major role for the data assimilation as long as the observations can be split into individual periods with a broken point at the collision instant.

	One future direction is to build cheap stochastic models to approximate the contact forces of the floe-floe interactions, which allow the development of Lagrangian data assimilation algorithms for the regions with a higher concentration of  sea ice floes. It is necessary to explore whether including the contact force will improve or deteriorate the data assimilation skill since the uncertainty in collision forces can be large. Another future work is to adopt a more sophisticated atmospheric model, including precipitation \cite{hu2021initial} as the forecast model, which allows to study the data assimilation skill in more refined regions.
	
	\section*{Acknowledgement}
	N.C. and G.M. are partially funded by ONR MURI N00014-19-1-2421. S.F. is a postdoc research associate under this grant. The authors thank Dr. Monica Martinez for providing the data of Figure \ref{Satellite_Image}.
	
	\section*{Appendix}
	
	\subsection{Details of the sea ice floe model}
	Recall the sea ice floe model \eqref{eqfirst}, where the total force $\bf F$ in \eqref{floe_force} has four components. The main text includes the description of the drag force from the atmosphere $\bf F_{\text{atm}}$ and that from the ocean $\bf F_{\text{ocn}}$. Here, we first discuss the other two forcing components.
	The force ${\bf F_{\text{pres}}}=(F_{\text{pres}}^x,F_{\text{pres}}^y) $ induced by the pressure of the ice is defined by	
	\begin{equation}\label{eqfpre}
	\begin{aligned}
	F_{\text{pres}}^x=&-\rho_{\text{ice}}h f_c{\bf v}^y_{\text{ocn}}\\
	F_{\text{prse}}^y=&\rho_{\text{ice}}h f_c{\bf v}^x_{\text{ocn}}
	\end{aligned}
	\end{equation}
	where $f_c$ is Coriolis parameter, $h$ is the thickness of the ice, and the vector ${\bf v}_{\text{ocn}}=({\bf v}^x_{\text{ocn}}, {\bf v}^y_{\text{ocn}})$ is the velocity of the ocean flow field. On the other hand, the Coriolis force ${\bf F}_{\text{cor}}=( F^x_{\text{cor}}, F^y_{\text{cor}})$ is defined by
	\begin{equation}
	\begin{aligned}
	F_{\text{cor}}^x=&\rho_{\text{ice}}h f_c{\bf v}^y_{\text{ice}}\\
	F_{\text{cor}}^y=&-\rho_{\text{ice}}h f_c{\bf v}^x_{\text{ice}}
	\end{aligned}
	\end{equation}
	
	Now we discuss the numerical calculation of the integral in Eq.\eqref{eqfirst}. We decompose the entire area of a floe into many small squares with equal sizes, which are parameterized by the radius $r$ and the angle $\psi$,
	\begin{equation}\label{eq2}
	\begin{split}
	m\frac{\d\bf{v}_{\text{cen}}}{\d t}=& \iint_{A}{\bf F} dA=\iint_{{(r,\psi)}}{\bf F}{(r,\psi)}dA,\\
	I\frac{\d\omega}{\d t}=&\iint_{A}\tau dA=\iint_{{(r,\psi)}}\tau{(r,\psi)}dA.
	\end{split}
	\end{equation}
	Note that ${\bf F}_{\text{ocn}}, {\bf F}_{\text{pres}}, \tau, {\bf v}_{\text{ocn}}$ and ${\bf v}_{\text{ice}} $ are functions of $(r, \psi)$, while ${\bf F}_{\text{w}}$ and ${\bf F}_{\text{cor}}$ are constants for each single floe. With these notations, the relationship of the floe velocity at point $(r,\psi)$ (assuming the center of the ice floe is at the origin) in the polar coordinate system can be defined as
	\begin{equation}\label{eqvice}
	\begin{aligned}
	{\bf v}_{\text{ice}}^x(r,\psi)=&{\bf v}_{\text{cen}}^x-r\Omega\sin(\psi),\\
	{\bf v}_{\text{ice}}^y(r,\psi)=&{\bf v}_{\text{cen}}^y+r\Omega\cos(\psi).
	\end{aligned}
	\end{equation}
	Similarly, the torque force in \eqref{eqt} can be computed via
	\begin{equation}
	\tau{(r,\psi)}=-r{\bf F}^x_{\text{io}}{(r,\psi)}\sin\psi+r{\bf F}^y_{\text{io}}{(r,\psi)}\cos\psi.
	\end{equation}
	Next, define the difference of ocean velocity and ice velocity as
	\begin{equation}\label{eqdv}
	\begin{aligned}
	dv_x=&{\bf v}^x_{\text{ocn}}-{\bf v}^x_{\text{ice}}\\
	dv_y=&{\bf v}^y_{\text{ocn}}-{\bf v}^y_{\text{ice}}
	\end{aligned}
	\end{equation}
	The force  ${\bf F}_{\text{ocn}}=( { F}_{\text{ocn}}^x, { F}_{\text{ocn}}^y ) $ induced by the sea-ice drag can be rewritten as
	\begin{equation}\label{eqfocn}
	\begin{aligned}
	F_{\text{ocn}}^x=&\rho_{\text{ocn}}c_{\text{ocn}} \sqrt{dv^2_x+dv^2_y} (\cos\theta dv_x+\sin\theta dv_y)\\
	F_{\text{ocn}}^y=&\rho_{\text{ocn}}c_{\text{ocn}} \sqrt{dv^2_x+dv^2_y}(-\sin\theta dv_x+\cos\theta dv_y)
	\end{aligned}
	\end{equation}
	where $\theta$ is a predefined turning angle of the ocean.

	\subsection{Proofs of the propositions}
	\subsubsection{Proof of Proposition \ref{prop:ou}}
	\begin{proof}
		Recall that for a linear process $\hat{u}_{\mathbf{k}}$ with Gaussian noise \eqref{eq:ocnmode}, the corresponding values of the variance and of the decorrelation time are given
		\begin{equation}
		\text{Var}(\hat{u}_{\mathbf{k}})=\frac{\sigma^2_{\mathbf{k}}}{2\gamma_{\mathbf{k}}}\qquad\mbox{and}\qquad
		\int_{0}^{\infty}R_{\bar{u}_{\mathbf{k}}}(\tau)d\tau=\frac{1}{\gamma_{\mathbf{k}}+i\omega_{\mathbf{k}}}.
		\end{equation}
		To find the unknown parameters, we solve
		\begin{equation}
		\frac{1}{d_{\mathbf{k}}+i\omega_{\mathbf{k}}}=T_{\mathbf{k}}-i\theta_{\mathbf{k}}\qquad\mbox{and}\qquad
		\frac{\sigma^2_{\mathbf{k}}}{2d_{\mathbf{k}}}=E_{\mathbf{k}},
		\end{equation}
		which yields \eqref{OU_Parameters}.
	\end{proof}
	
	\subsubsection{Proof of Proposition \ref{Prop_nonG1}}
	\begin{proof}
		The stationary PDF $p(\hat{u}_{\mathbf{k}, 1},\hat{u}_{\mathbf{k}, 2})$ associated with the system \eqref{2D_model} satisfies the Fokker-Planck equation,
		\begin{equation}
		\frac{\partial^2}{\partial \hat{u}_{\mathbf{k}, 1}^2}\Big(\sigma_{\mathbf{k},1}^2p\Big) + \frac{\partial^2}{\partial \hat{u}_{\mathbf{k}, 2}^2}\Big(\sigma_{\mathbf{k},2}^2p\Big) = 2\frac{\partial}{\partial \hat{u}_{\mathbf{k}, 1}}\Big((-\gamma_{\mathbf{k}}, \hat{u}_{\mathbf{k}, 1} - \omega_{\mathbf{k}} \hat{u}_{\mathbf{k}, 2})p\Big) + 2\frac{\partial}{\partial \hat{u}_{\mathbf{k}, 2}}\Big((\omega_{\mathbf{k}} \hat{u}_{\mathbf{k}, 1} - \gamma_{\mathbf{k}} \hat{u}_{\mathbf{k}, 2})p\Big).
		\end{equation}
		To find one solution of $\sigma_1(u_1,u_2)$ and $\sigma_2(u_1,u_2)$, it is natural to let
		\begin{equation}\label{split_equation}
		\begin{split}
		\frac{\partial^2}{\partial u_{\mathbf{k}, 1}^2}\Big(\sigma_{\mathbf{k},1}^2p\Big) &= 2\frac{\partial}{\partial \hat{u}_{\mathbf{k}, 1}}\Big((-\gamma_{\mathbf{k}} \hat{u}_{\mathbf{k}, 1} - \omega_{\mathbf{k}} \hat{u}_{\mathbf{k}, 2})p\Big),\\
		\frac{\partial^2}{\partial u_{\mathbf{k}, 2}^2}\Big(\sigma_{\mathbf{k},2}^2p\Big) &= 2\frac{\partial}{\partial \hat{u}_{\mathbf{k}, 2}}\Big((\omega_{\mathbf{k}} \hat{u}_{\mathbf{k}, 1} - \gamma_{\mathbf{k}} \hat{u}_{\mathbf{k}, 2})p\Big).
		\end{split}
		\end{equation}
		Taking twice the integration with respect to $u_1$ and $u_2$ for the first and second equations in \eqref{split_equation}, respectively, yields
		\begin{equation}\label{split_equation2}
		\begin{split}
		\sigma_1^2(\hat{u}_{\mathbf{k}, 1},\hat{u}_{\mathbf{k}, 2})p(\hat{u}_{\mathbf{k}, 1},\hat{u}_{\mathbf{k}, 2}) &= 2\int_{-\infty}^{u_1}(-\gamma_{\mathbf{k}} s - \omega_{\mathbf{k}} \hat{u}_{\mathbf{k}, 2})p(s,\hat{u}_{\mathbf{k}, 2})\d s,\\
		\sigma_2^2(\hat{u}_{\mathbf{k}, 1},\hat{u}_{\mathbf{k}, 2})p(\hat{u}_{\mathbf{k}, 1},\hat{u}_{\mathbf{k}, 2}) &= 2\int_{-\infty}^{u_2}(\omega_{\mathbf{k}} \hat{u}_{\mathbf{k}, 1} - \gamma_{\mathbf{k}} s)p(\hat{u}_{\mathbf{k}, 1},s)\d s.
		\end{split}
		\end{equation}
		Dividing both side by $p(\hat{u}_{\mathbf{k}, 1},\hat{u}_{\mathbf{k}, 2})$ leads to \eqref{Multiplicative_noise_2states}.
	\end{proof}
	
	\subsubsection{Proof of Proposition \ref{Prop_nonG2}}
	\begin{proof}
		Starting from \eqref{split_equation2}, where now we formally assume $\sigma_{\mathbf{k},1}:=\sigma_{\mathbf{k},1}(\hat{u}_{\mathbf{k},1})$ and $\sigma_{\mathbf{k},2}:=\sigma_{\mathbf{k},2}(\hat{u}_{\mathbf{k},2})$,
		\begin{subequations}\label{split_equation3}
			\begin{align}
			\sigma^2_{\mathbf{k},1}(\hat{u}_{\mathbf{k},1})p(\hat{u}_{\mathbf{k},1},\hat{u}_{\mathbf{k},2}) &= 2\int_{-\infty}^{\hat{u}_{\mathbf{k},1}}(-\gamma_{\mathbf{k}} s - \omega_{\mathbf{k}} \hat{u}_{\mathbf{k},2})p(s,\hat{u}_{\mathbf{k},2})\d s,\label{split_equation3_1}\\
			\sigma^2_{\mathbf{k},2}(\hat{u}_{\mathbf{k},2})p(\hat{u}_{\mathbf{k},1},\hat{u}_{\mathbf{k},2}) &= 2\int_{-\infty}^{\hat{u}_{\mathbf{k},2}}(\omega_{\mathbf{k}} \hat{u}_{\mathbf{k},1} - \gamma_{\mathbf{k}} s)p(\hat{u}_{\mathbf{k},1},s)\d s.\label{split_equation3_2}
			\end{align}
		\end{subequations}
		It is important to note that the equalities in \eqref{split_equation3} may not be valid since the right hand side of \eqref{Multiplicative_noise_2states_1} is in general a function of both $\hat{u}_{\mathbf{k},1}$ and $\hat{u}_{\mathbf{k},2}$ while $\sigma_{\mathbf{k},1}$ is only a function of $\hat{u}_{\mathbf{k},1}$ (similar argument for \eqref{Multiplicative_noise_2states_2} and $\sigma_{\mathbf{k},2}$). Therefore, an integration with respect to $u_2$ and $u_1$ is taken for both sides of \eqref{split_equation3_1} and \eqref{split_equation3_2}, respectively. The solutions of $\sigma_{\mathbf{k},1}(\hat{u}_{\mathbf{k},1})$ and $\sigma_{\mathbf{k},2}(\hat{u}_{\mathbf{k},2})$ are searched based on such averaged equations. For $\sigma_{\mathbf{k},1}(\hat{u}_{\mathbf{k},1})$, the solution is given by
		\begin{equation}\label{split_equation4}
		\begin{split}
		\sigma^2_{\mathbf{k},1}(\hat{u}_{\mathbf{k},1})p(\hat{u}_{\mathbf{k},1}) &= 2\int_{-\infty}^{\infty}\int_{-\infty}^{\hat{u}_{\mathbf{k},1}}( \gamma_{\mathbf{k}} s - \omega_{\mathbf{k}} \hat{u}_{\mathbf{k},2})p(s,\hat{u}_{\mathbf{k},2})\d s\d \hat{u}_{\mathbf{k},2},\\
		&=-2\gamma_{\mathbf{k}} \int_{-\infty}^{\hat{u}_{\mathbf{k},1}}s\left(\int_{-\infty}^{\infty}p(s,\hat{u}_{\mathbf{k},2})d\hat{u}_{\mathbf{k},2}\right) \d s - 2\omega_{\mathbf{k}}\int_{-\infty}^{u_1} \left(\int_{-\infty}^{\infty} u_2p(s,\hat{u}_{\mathbf{k},2})d\hat{u}_{\mathbf{k},2}\right)\d s,\\
		&=-2\gamma_{\mathbf{k}} \int_{-\infty}^{\hat{u}_{\mathbf{k},1}}sp_1(s)\d s,
		\end{split}
		\end{equation}
		where $\int_{-\infty}^{\infty} \hat{u}_{\mathbf{k},2}p(s,\hat{u}_{\mathbf{k},2})\d\hat{u}_{\mathbf{k},2}=0$ for all $s$. This is because $\hat{u}_{\mathbf{k},1}$ and $\hat{u}_{\mathbf{k},2}$ are the real and imaginary parts of one Fourier coefficients which are orthogonal to each other. In addition, the long-term mean of $\hat{u}_{\mathbf{k},1}$ and $\hat{u}_{\mathbf{k},2}$ are zero in the absence of constant forcings in \eqref{2D_model}.
	\end{proof}
	\bibliography{references}
\end{document}